\documentclass{ws-mpla}
\usepackage[super]{cite}
\usepackage{graphicx}
\usepackage{dcolumn}
\usepackage{bm}
\usepackage{epsfig,amsmath}
\usepackage[usenames,dvipsnames]{color}
\usepackage{appendix}
\usepackage{textcomp}
\usepackage{float}
\usepackage{subfigure}
\begin{document}


\catchline{}{}{}{}{}

\title{$P-V$ Criticality of Conformal Gravity Holography in Four Dimensions}

\author{Parthapratim Pradhan}

\address{Department of Physics, Hiralal Mazumdar Memorial College For Women
\footnote{On Lien from the Department of Physics, Vivekananda Satavarshiki Mahavidyalaya,
Manikpara, Jhargram, West Bengal~721513, India}, Dakshineswar, Kolkata-700035, 
West Bengal, India}

\maketitle

\begin{abstract}
{We examine the critical behaviour i. e. $P-V$ criticality  of conformal gravity~(CG) 
 in  an extended phase space in which the cosmological 
constant should be interpreted as a thermodynamic pressure and the corresponding  conjugate quantity 
as a thermodynamic volume.}
The main potential point of interest in CG is that there exists a {non-trivial} \emph{Rindler 
parameter ($a$)} in the {spacetime geometry. This geometric parameter has an important role to 
construct a model for gravity at large distances where the parameter ``$a$'' actually originates }.  
We also investigate the effect  of the said parameter on the {black hole~(BH)  
\emph{thermodynamic} equation of state, critical constants, Reverse Isoperimetric Inequality,} 
{first law of thermodynamics, Hawking-Page phase 
transition and Gibbs free energy} for this BH.  We speculate that 
due to the presence of the said parameter,  there has been a deformation {in the shape}  
of {the} isotherms in the $P-V$ diagram in comparison with {the} 
charged-AdS~(anti de-Sitter) BH and {the} chargeless-AdS BH. 
Interestingly, we find {that} the \emph{critical ratio} for this BH is 
$\rho_{c} = \frac{P_{c} v_{c}}{T_{c}}= \frac{\sqrt{3}}{2}\left(3\sqrt{2}-2\sqrt{3}\right)$, which is  
greater than the charged AdS BH and Schwarzschild-AdS BH {i.e.}
$\rho_{c}^{CG}:\rho_{c}^{Sch-AdS}:\rho_{c}^{RN-AdS} = 0.67:0.50:0.37$.  The symbols are defined in the 
main work. Moreover, we observe that \emph{{the} critical ratio 
{has a constant value}} and {it is} independent of the {non-trivial} \emph{Rindler 
parameter ($a$)}. 
Finally, we derive {the} \emph{reduced equation of state} in terms of {the}
\emph{reduced temperature}, {the} \emph{reduced volume} and {the}
\emph{reduced pressure} respectively.

\end{abstract}

.

\section{Introduction}
The study of thermodynamic properties of BH in the AdS space has gained much more 
attention in recent years due to the {seminal} work of Hawking and Page~\cite{haw83}. 
The AdS case is particularly interesting because of gauge-gravity duality via dual conformal 
field theory~(CFT). {This duality indicates that the spherically symmetric charged AdS BH 
admits critical behaviour which is similar to the liquid-gas phase transition. It was explicitly 
described  by Chamblin et al.~\cite{chamblin99,chamblin99a,emparan}.} 
They {proved that} there should {exist} the first order phase transition 
{in the conventional phase space} in case of Reissner-Nordstr\"{o}m-AdS~(RN-AdS) BH. 
The critical behaviour of this BH has been studied there in details. Furthermore they emphasized 
{that} this behaviour is quite analogous to the Van-der-Waal's~(VdW's) 
{type liquid-gas phase transition}.

{Now if one could treat} the cosmological constant as a  thermodynamic 
pressure~\cite{kastor09,dolan10,dolan11,cvetic11}, the Arnowitt-Deser-Misner~(ADM) mass of the AdS BH as 
{an} enthalpy of the thermodynamic system and the thermodynamically conjugate 
quantity {as} a thermodynamic volume then one {could} study 
the {critical} behaviour in the extended phase space. {This is  
an active area of research over the past few years.}

{Moreover, one could} investigate the critical { behaviour in the 
extended phase space} which is {quite} analogous 
to VdW's {like} {liquid-gas 
phase transition}. This thermodynamic  {behaviour} for RN-AdS BH has 
been {explicitly} studied  by 
Kubiz\v{n}\'{a}k-Mann~\cite{david12} by using the extended phase-space formalism. They {postulated} 
the analogy between VdW's fluid-gas system and charged AdS BH. They also derived  the
BH equation of state, and determined the critical constants in comparison with the liquid-gas system. 
The critical constants {and} critical exponents were calculated {therein}.
{It has been shown that they coincide}  with those of the liquid-gas systems. 
The critical behaviour { for different types of BH has been studied elaborately in the 
extended phase space and the same could} be found in Ref.~\cite{chen,li}.

In {the present study, we wish}  to examine {the critical behaviour i.e.}
$P-V$ criticality of CG holography~\cite{grum14} in four dimensions~(4D) by treating the cosmological 
{parameter} as the thermodynamic {pressure} and 
its conjugate {parameter} as thermodynamic volume. We derive 
the BH equation of state in terms of BH temperature and specific 
thermodynamic volume. At the critical point, we calculate the critical constants. It {has 
been }shown that the critical ratio {has a constant value} which is quite different from RN-AdS BH.  
We {would} also derive the first law of BH thermodynamics, 
{Reverse Isoperimetric Inequality} and Gibbs free energy. {Moreover, we would 
discuss the Hawking-Page phase transition. It has been shown that the Hawking-Page phase transition temperature 
as a function of Rindler parameter.} Finally, we {would} derive the reduced 
equation of state. 

Furthermore,  we {would prove} that the three critical constants namely the critical 
pressure, the critical temperature and the critical volume {do all} 
\emph{depend on the Rindler parameter}. The BH {thermodynamic} equation of state 
also depends on the said parameter. Interestingly, the critical ratio is \emph{independendent of the Rindler parameter}. 
Due to  {the presence of} this parameter the shape of the isotherms 
{in the} $P-V$ diagram {of}
CG  {BH} is completely different from  { that of the} Schwarzschild-AdS spacetime 
and  {the} RN-AdS space-time. 
This is one of the interesting {observations} of this work.

The {CG} is a fascinating theory of gravity {and} has a \emph{non-trivial Rindler term}. 
{This subleading term indeed originates during the formulation of an effective model 
for gravity at large distances~\cite{grum10}.}
{This novel parameter produces an acceleration which is {the already known} 
\emph{Rindler acceleration~ ($a$). 
For example, it produces an anomolous acceleration in geodesics of a test particle.~\cite{grum10}}. 
It could be observed in various ``anomolous'' systems like galaxy-cluster, star-galaxy,  Earth-satellite etc.  
In Ref.~\cite{grum10}, the author described an effective 2D field theory for infra-red~(IR) gravity to explain 
the anomolous acceleration of a test particle in the gravitational field of a central object. He also derived
an action\footnote{The equation of motion, the line element and two dimensional Ricci scalar for this IR gravity 
see the Ref.~\cite{grum10}~(for details).} for this gravity 
in the IR region as 
\begin{eqnarray}
{\mathcal {I}} &=&-\int {\sqrt {-g}} \, {d}^{2}x \, 
\left(\Phi^2R+2(\partial \Phi)^2-6\Lambda\Phi^2+8a\Phi+2 \right) ~.\label{ir}
\end{eqnarray}
where $\Phi$ is a scalar field, $\Lambda$ is a cosmological constant and $a$ is Rindler term which generates the 
Rindler acceleration in a geodesics. This acceleration also depends on the scale of the system that one may use.
The value of $a$ in the Ref.~\cite{grum10} has been given $a\approx 10^{-62}-10^{-61}$ for anomolous 
acceleration. For Pioneer acceleration, the value of $a$ is approximately $a\approx 10^{-61}$
whereas for modified-Newtonian-dynamics acceleration the value of $a$ is around $a\approx 10^{-62}$.
The main motivation behind this work is to investigate the effect of the Rindler acceleration
on the thermodynamic properties of the CG holography in four dimensions. }

{The other aspect is that} CG is a theory of gravity at large distances~\cite{grum10}. Like other 
higher curvature theory, it is also a re-normalizable theory having ghost ~\cite{stele,adler}. 
On the other hand, the Einstein's general theory of gravity has no ghost i.e. ghost 
free gravity but two loop non-renormalizable~\cite{goroff}. To explain galactic rotating curves without 
dark matter, Mannheim was first { to study this theory} phenomenologically ~\cite{mann}. 
It {emerges} as a counter term in AdS/CFT~(conformal field theory) correspondence~ \cite{liu,bala}. 

In the  quantum gravity context, CG has been studied by 't Hooft~ \cite{hooft}. Maldacena~\cite{maldacena} 
showed that by imposing appropriate 
boundary  condition it is possible to eliminate the ghost term. The most important feature of this theory is that it depends 
only on the~(Lorentz) angles but not on the distance. It should be noted that CG is a higher-derivative theory but
the entropy  obeys the area law ~\cite{grum14}.   {Another striking} feature of CG is that the  
AdS boundary condition is weaker than the Starobinsky boundary {conditions}~\cite{star}. 
{This may admit a non-trivial asymptotic Rindler term concurrently with Grumiler's proposal 
`an effective model for gravity at large distances'  given  in 2010~\cite{grum10}}.
Thus one {could} observe a new critical behaviour of CG BH compared to 
Einstein's gravity due to the non-trivial {geometrical} Rindler parameter.

The {structure} of the paper is as follows. In Sec.~(\ref{cgg}), we have described the thermodynamic 
properties of CG holography in four dimensions. Finally, we have given {the} conclusions in 
Sec.~(\ref{dis}).
\section{\label{cgg} Thermodynamic  {Properties}  of CG Holography in Four Dimensions}
{Before begining the main investigation, we would like to review some basic features of CG. 
It is a theory derived from the action of the Weyl conformal curvature. The Weyl action concurrently with 
an electromagnetic field is given by~\cite{rigert84} 
\begin{eqnarray}
{\mathcal {I}} &=&-\frac{1}{4} \int {\sqrt {-g}} \, {d}^{4}x \, 
\left(\beta C_{abcd}C^{abcd} +F^{cd}F_{cd} \right) ~.\label{ac}
\end{eqnarray}
where $C_{abcd}$ is denoted as the conformal curvature tensor and $F_{ab}\equiv A_{a,b}-A_{b,a}$ is denoted 
as electromagnetic field tensor. Using Eq.~(\ref{ac}), one could write the Bach-Maxwell equations by varying 
of $g_{ab}$ and $A_{b}$ as 
\begin{eqnarray}
\left(\nabla^{a}\nabla^{b}-\frac{1}{2}R^{ab} \right) C_{acbd}  &=& \frac{1}{2\beta} 
\left( F_{c}^{\xi}F_{d\xi} -\frac{1}{4} g_{cd}F^{ab}F_{ab}  \right) ~,\label{ac1}\\
\nabla^{a}F_{ab} &=& 0
\end{eqnarray}
Therefore the most static, spherically symmetric solution of Eq.~(\ref{ac1}) can be written as~\cite{rigert84} 
\begin{eqnarray}
ds^2 &= & -{\cal F}(r) dt^2 + \frac{dr^2}{{\cal F}(r)} +r^2 \left(d\theta^2+\sin^2\theta d\phi^2\right) .~\label{mtcg}
\end{eqnarray}
where the meric function has the general form given by 
\begin{eqnarray}
{\cal F}(r) &=&  \lambda r^2+\mu r+\delta+\frac{\eta}{r} .~\label{h2}
\end{eqnarray}
and 
\begin{eqnarray}
 A \equiv A_{a}dx^a=\frac{q}{r} dt
\end{eqnarray}
where $\lambda$, $\mu$, $\delta$, $\eta$ are constants satisfied by the following relation~\cite{rigert84}
\begin{eqnarray}
 3\mu \eta-\delta^2+1+\frac{3q^2}{2\beta} &=& 0 .~\label{h3}
\end{eqnarray}
It should be noted that the spherically symmetric solutions of CG were first introduced by Bach~\cite{bach21} in 1921. 
Later Buchdahl~(1953)~\cite{buch53} considered a particular case $\mu=q=0$, $\delta=1$ by using Eq.~(\ref{h2}).}

{In 2010, Grumiler~\cite{grum10} proposed an effective model for gravity of a central object at a large 
distance. For the large radius expansions he found a solution which consists of a cosmological constant and an 
additional parameter which is called the ``Rindler parameter''. This novel parameter could produce acceleration analogous 
to that observed in different `anomolous~\footnote{The meaning of this term is that ``the difference between 
the observed trajectory of a test particle in the gravitational field of a central object and the calculated 
trajectory''.} systems~(i.e. Sun-Pioneer spacecraft, star-galaxy etc.).}  

{Subsequently in 2014, the same author gave a spherically symmetric 
solution of CG  in another form by substituting 
the values of $\lambda$, $\mu$, $\delta$, $\eta$ in Eq.~(\ref{h2}) of Ref.~\cite{rigert84} as 
$\lambda=-\frac{\Lambda}{3}$, $\mu=2a$, $\delta=\sqrt{1-12am}$ and $\eta=-2M$ and without 
considering the Maxwell field i.e.  keeping $q=0$. Interestingly, the values of these parameters 
satisfied the Eq.~(\ref{h3}). Therefore the form of metric function can be written as  
}
\begin{eqnarray}
{\cal F}(r) &=&  \sqrt{1-12aM}-\frac{2M}{r}+2ar-\frac{\Lambda}{3} r^2 .~\label{h4}
\end{eqnarray}
Where $a$ is the Rindler parameter {and $a\geq 0$}. In the limit $a=0$, one finds 
the Schwarzschild-AdS space-time. In the limit $aM<<1$, one obtains {the} Grumiler 
space-time~\cite{grum10}. {When we set the parameters $a=\Lambda=0$, we obtain the 
Schwarzschild solutions, with $M$ being the BH mass. The numerical values of $\Lambda$, $a$ 
and $M$ (in Planck units) are considered in~\cite{grum14} as $\Lambda \approx 10^{-123}$, 
$a \approx 10^{-61}$ and $M \approx 10^{38} M_{\star}$, where $M_{\star}=1$ for the sun, and therefore 
it implies $aM \approx 10^{-23} M_{\star} \ll 1$ for all types of BHs or galaxies in our Universe. 
}

Let us now put $-\frac{\Lambda}{3}=\frac{1}{\ell^2}$ for  {the} AdS case.
The BH event horizon, $r_{+}$ can be obtained by {imposing the conditon}
${\cal F}(r_{+})=0$ i.e. 
\begin{eqnarray}
 r_{+}^3 +2a\ell^2r_{+}^2+\sqrt{1-12aM}\ell^2r_{+}-2M\ell^2 &=& 0 ~.\label{eq1}
\end{eqnarray}

\begin{figure}[h]
  \begin{center}
\subfigure[]{
\includegraphics[width=2.1in,angle=0]{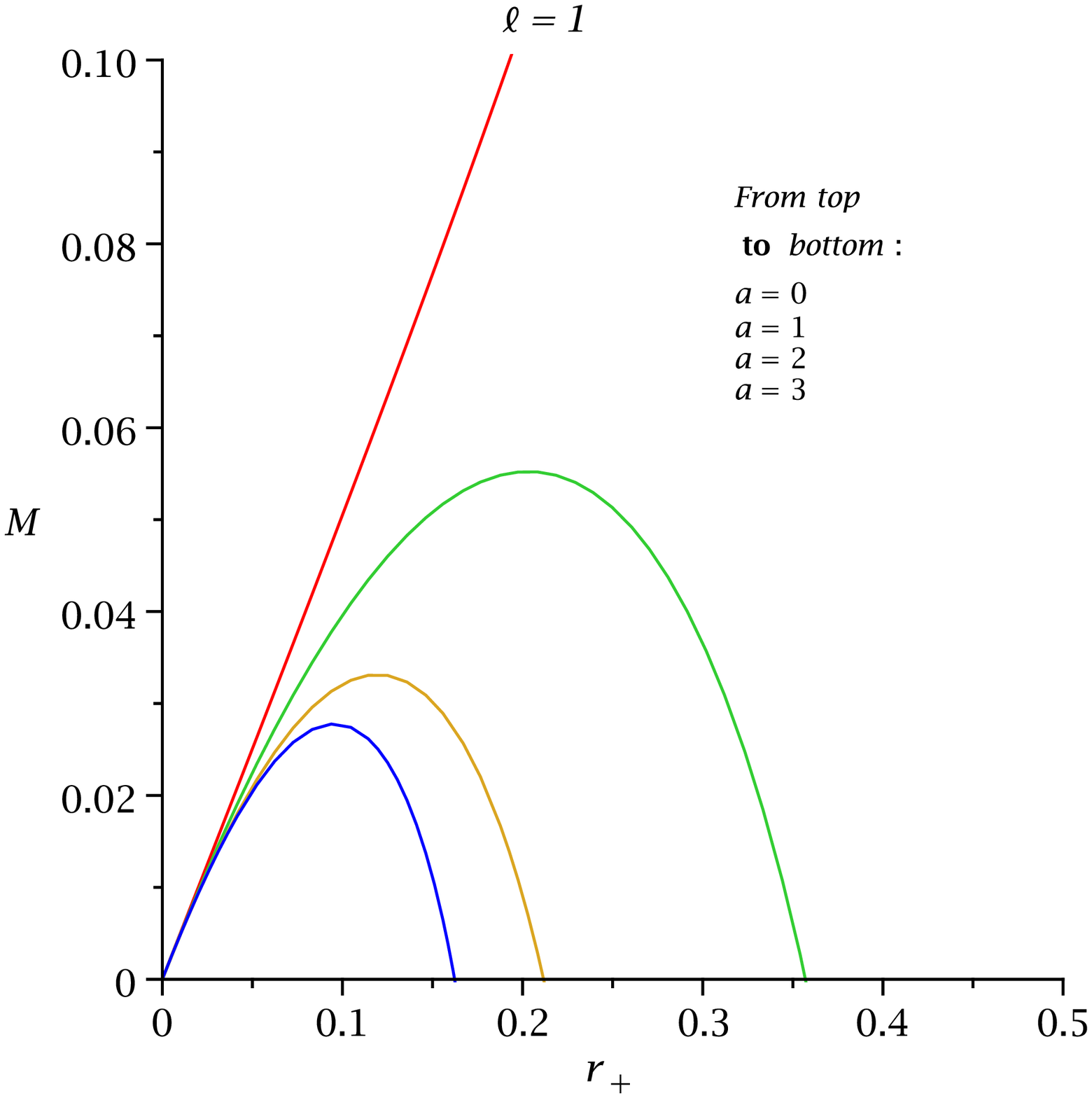}} 
 \subfigure[]{
 \includegraphics[width=2.1in,angle=0]{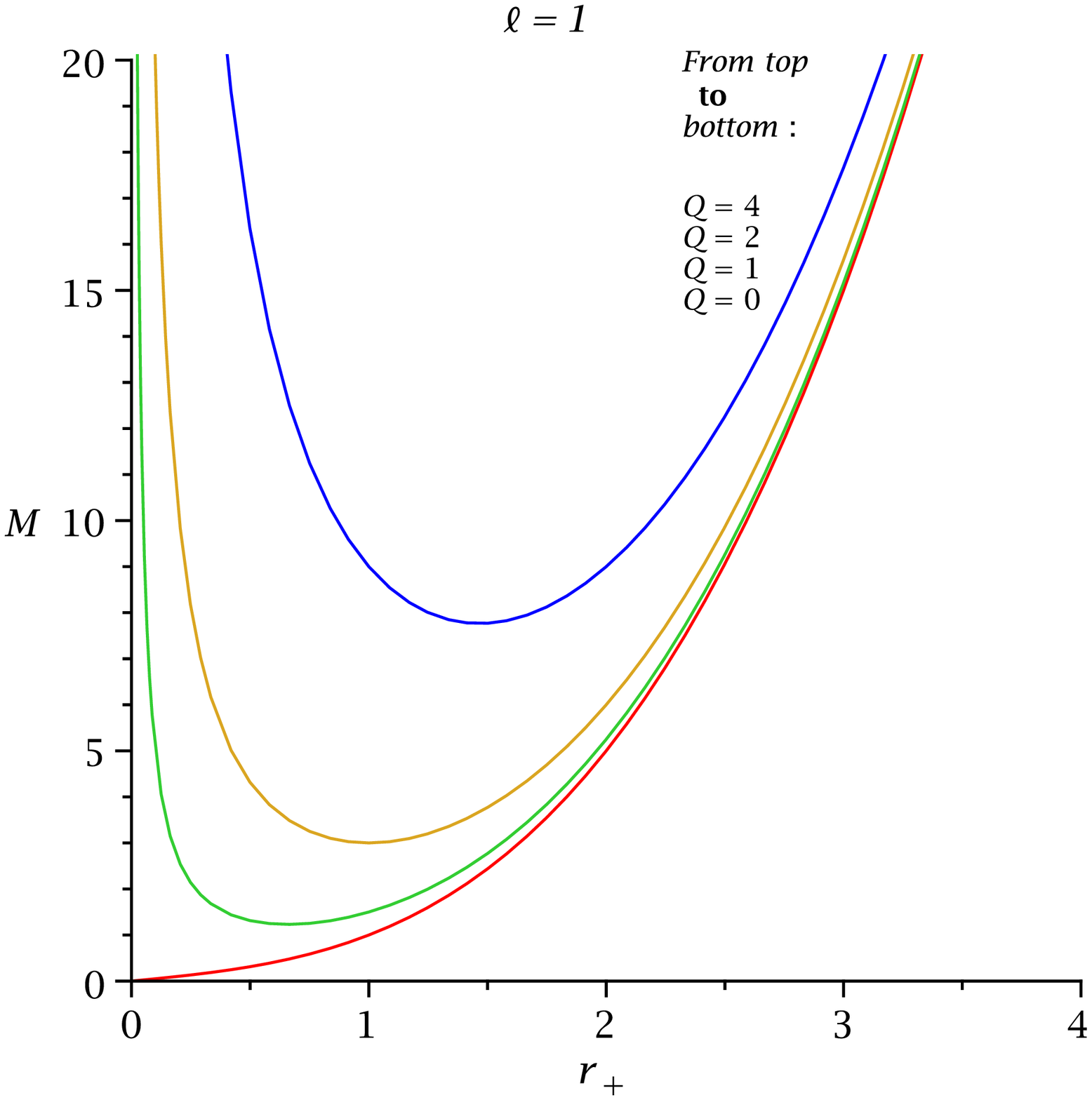}}
 \caption{\label{mf}\textit{Variation  of $M$  with $r_{+}$ for Schwarzschild-AdS BH, CG BH and 
 RN-AdS space-time.}}
\end{center}
\end{figure}

\begin{figure}[h]
  \begin{center}
\subfigure[]{
\includegraphics[width=2.5in,angle=0]{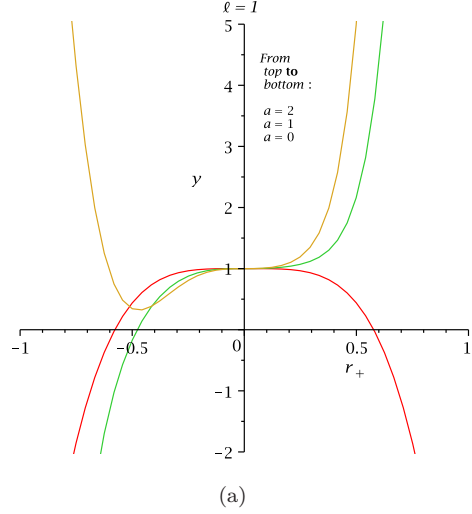}} 
 \caption{\label{yf}\textit{Variation  of $y=54 a r_{+}^7+36 a^2 \ell^2 r_{+}^6-9 \ell^2 r_{+}^4 +6 a \ell^4 r_{+}^3 +\ell^6$  
 with $r_{+}$ for CG  BH. For $a=0$, the red one curve implies the instability, whereas for $a=1, 2$: the green one 
 and the yellow one signals a stability.}}
\end{center}
\end{figure}

\begin{figure}[h]
  \begin{center}
\subfigure[]{
\includegraphics[width=2.1in,angle=0]{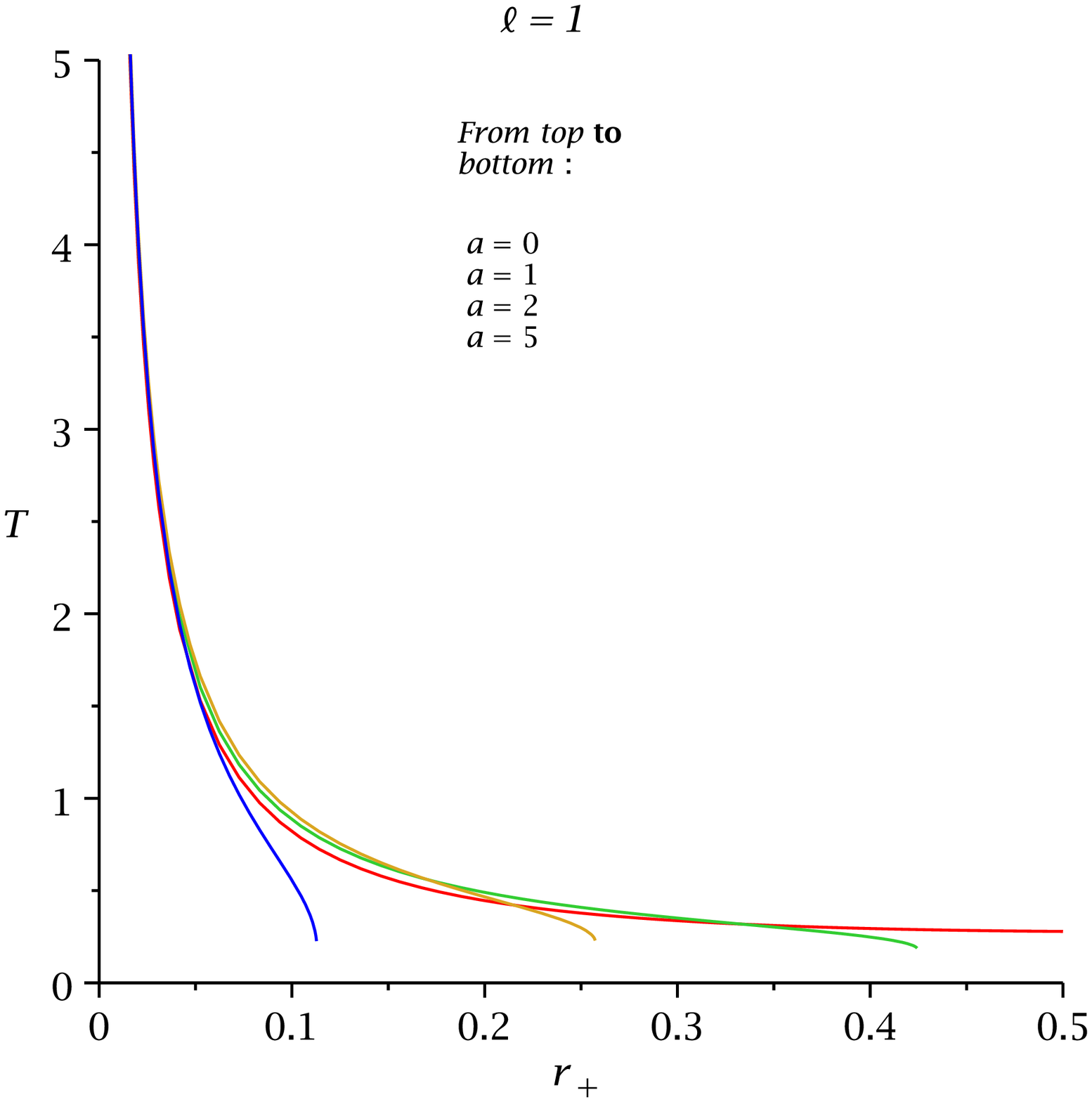}} 
 \subfigure[]{
 \includegraphics[width=2.1in,angle=0]{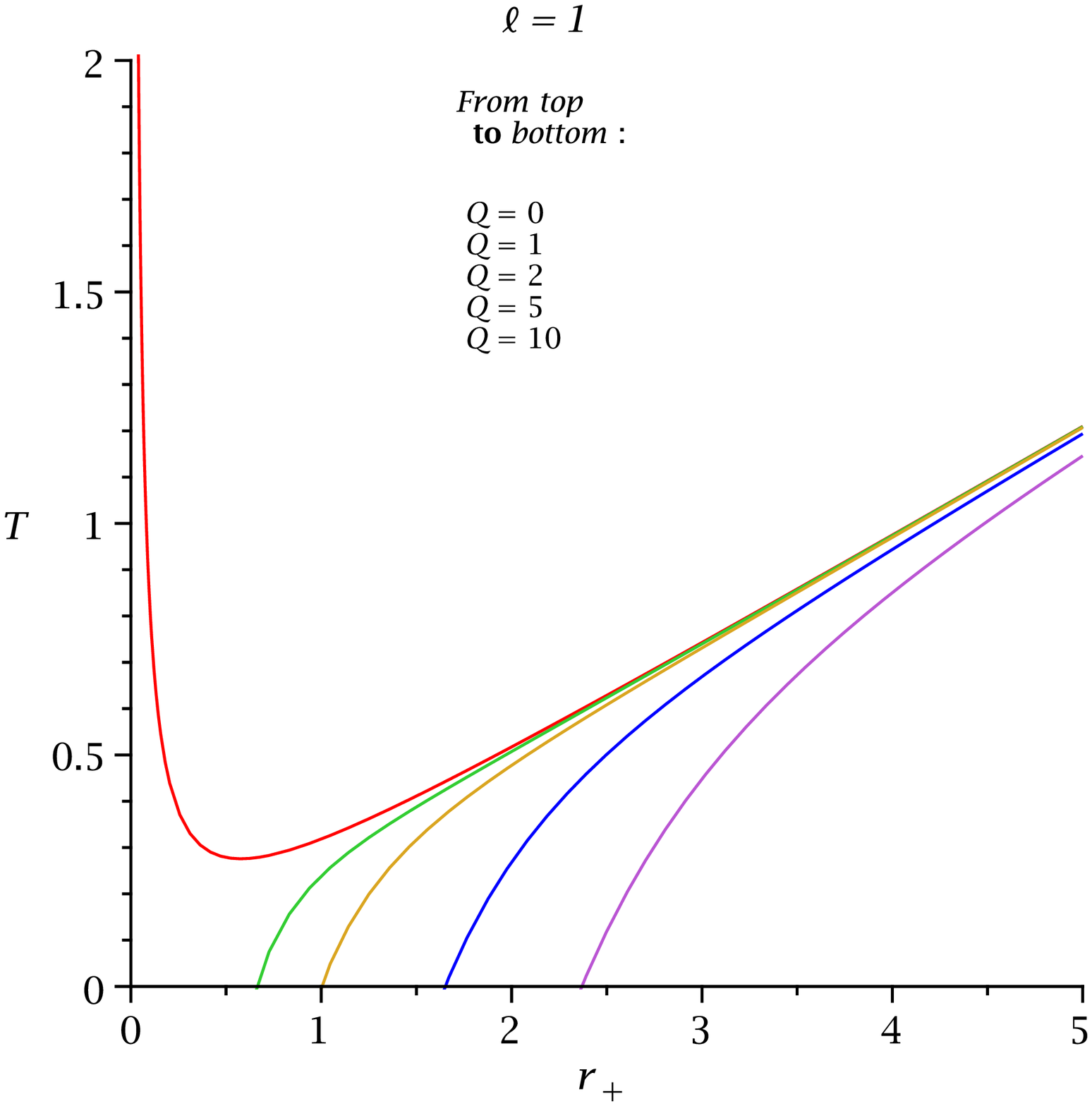}}
 \caption{\label{tf}\textit{Variation  of $T$  with $r_{+}$ for Schwarzschild-AdS BH, CG BH and RN-AdS BH. 
From the left figure, it follows that the parameter $a$ is modified the temperature {curve in comparison}
with the Schwarzschild-AdS BH. In case of charged AdS BH, it is completely different {in} shape.}}
\end{center}
\end{figure}

By solving the above equation, one obtains the mass parameter in terms of the 
event horizon radius as
\begin{eqnarray}
M  &=& \frac{r_{+}}{2}\left[\sqrt{1-3a^2r_{+}^2-\frac{6ar_{+}^3}{\ell^2}}-a r_{+}
+\frac{r_{+}^2}{\ell^2}\right] ~.\label{eq2}
\end{eqnarray}
In the limit $a=0$, we get the ADM mass for Schwarzschild-AdS BH. It is given by 
\begin{eqnarray}
M  &=& \frac{r_{+}}{2}\left[1+\frac{r_{+}^2}{\ell^2}\right] ~.\label{eqq2}
\end{eqnarray}
It indicates that the mass parameter is a function of event horizon radius and {is}
strictly increasing function. But for 
CG BH the mass parameter is a function of both the event horizon radius and the Rindler 
parameter. It seems that due to 
the Rindler acceleration the mass function first increases when the horizon radius 
increases then it decreases sharply. {It} is reverse in 
{nature due to } the presence of charge parameter and it could be observed 
from the Fig.~\ref{mf}. In the presence of charge parameter, the mass function becomes 
\begin{eqnarray}
M  &=& \frac{r_{+}}{2}\left[1+\frac{Q^2}{r_{+}^2}+\frac{r_{+}^2}{\ell^2}\right] ~.\label{eqq3}
\end{eqnarray}
The BH temperature~\cite{pp} is given by 
\begin{eqnarray}
T &=& 
\frac{1}{4\pi r_{+}} \left(\sqrt{1-3a^2r_{+}^2-\frac{6a}{\ell^2}r_{+}^3}+a r_{+}+3\frac{r_{+}^2}{\ell^2}\right)
~. \label{eq3}
\end{eqnarray}
{It indicates that the BH temperature is dependent on the Rindler parameter}. 
When $a=0$, one obtains the temperature of {the} famous Schwarzschild-AdS BH.
The maximum and minimum value of the above temperature could be found from the following condition
\begin{eqnarray}
\left(\frac{\partial T}{\partial r_{+}}\right)|_{a,\ell} &=& 0   ~.\label{tmn}
\end{eqnarray}
which gives 
\begin{eqnarray}
54 a r_{+}^7+36 a^2 \ell^2 r_{+}^6-9 \ell^2r_{+}^4 +6 a \ell^4 r_{+}^3 +\ell^6 &=& 0  ~.\label{tmn1}
\end{eqnarray}
{It is a non-trivial task to determine the roots of the $7^{th}$ order polynomial 
equation. Rather, it is easy to  see what happens if the limit is $a=0$}. {One obtains}
\begin{eqnarray}
 r_{+} &=& \frac{\ell}{\sqrt{3}} =\frac{1}{\sqrt{8 \pi P}}  ~.\label{tmn2}
\end{eqnarray}
where the temperature is $T=T_{min}=\frac{\sqrt{3}}{2\pi \ell}=\frac{0.275}{\ell}$ and a single BH is formed with 
horizon radius $r_{+}$. When $T<T_{min}$, there are no BHs { formed}. While for $T>T_{min}$, there 
exists a small {and a} large BH.  Their radii {could} be calculated from the 
Eq.~(\ref{eq3}) (in the limit $a=0$). We have plotted the Eq.~(\ref{tmn1}) in the Fig.~\ref{yf} to 
{distinguish the stable and the unstable region.} 
By introducing {the} charge parameter we get the temperature of RN-AdS BH. 
We do no write the explicit expressions but our aim is to show the variation of this 
temperature with {the} event horizon radius and compared it with our model. It 
could be {found} from the Fig.~\ref{tf}. 
The entropy~\cite{grum14} of the BH using Wald's formalism is 
\begin{eqnarray}
{\cal S}_{i}  &=& \frac{{\cal A}_{i}}{4\ell^2} ~.\label{eq5}
\end{eqnarray}
where the area of the BH is given by
\begin{eqnarray}
{\cal A}_{i}  &=& 4\pi r_{i}^2 ~.\label{eq6}
\end{eqnarray}
Interestingly, the entropy relation {fulfils} the area law 
despite the fact that CG is a higher derivative gravity.

Since in this work we are {interested to study} the $P-V$ criticality 
in the extended phase space therefore one {could} define the cosmological 
constant as thermodynamic pressure and {the corresponding} conjugate variable 
as thermodynamic volume {i.e.}
\begin{eqnarray}
P &=& -\frac{\Lambda}{8\pi}=\frac{3}{8\pi \ell^2} ~.\label{eq8}
\end{eqnarray}
and
\begin{eqnarray}
V &=& \left(\frac{\partial M}{\partial P}\right)_{S}  ~.\label{eq9}
\end{eqnarray}
Now in the extended phase space, the mass parameter becomes
\begin{eqnarray}
M  &=& \frac{r_{+}}{2}\left[\sqrt{1-3a^2r_{+}^2-16\pi a Pr_{+}^3}-a r_{+}
+\frac{8\pi P}{3}r_{+}^2\right] ~.\label{eq12}
\end{eqnarray}
{and the thermodynamic volume should be} 
\begin{eqnarray}
V &=& \left(\frac{\partial M}{\partial P}\right)_{S} 
=\frac{4}{3}\pi r_{+}^3\left[1-\frac{3ar_{+}}{\sqrt{1-3a^2r_{+}^2-16\pi Par_{+}^3}}\right] 
~.\label{cg1}
\end{eqnarray}
\begin{figure}[h]
  \begin{center}
\subfigure[]{
\includegraphics[width=2.5in,angle=0]{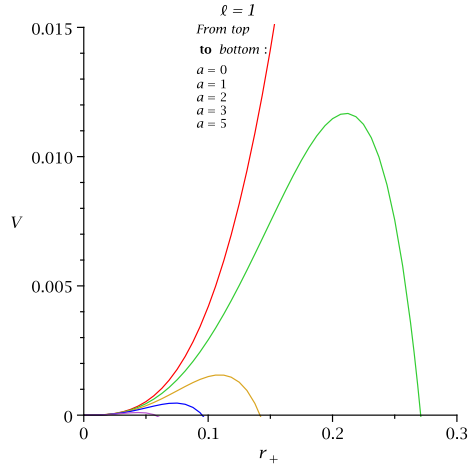}} 
 \caption{\label{vf}\textit{Variation  of $V$  with $r_{+}$ for Schwarzschild-AdS BH and CG  BH. }}
\end{center}
\end{figure}
It is quite strange that due to the \emph{Rindler acceleration} the thermodynamic volume {gets} 
modified and it has been shown in Fig.~\ref{vf}. It also indicates that the thermodynamic volume depends upon the
non-trivial Rindler parameter. This is probably \emph{ the first counter example} of any spherically symmetric BH 
{whose} thermodynamic volume is 
\begin{eqnarray}
V & \neq & \frac{4}{3}\pi r_{+}^3  ~.\label{eq10}
\end{eqnarray}
In general, we know that the thermodynamic volume~\cite{kastor09,dolan10,dolan11,cvetic11,mann1,david12} for 
spherically symmetric BH~(For example, RN-AdS BH ~\cite{david12}) 
is given by 
\begin{eqnarray}
V & = & \frac{4}{3}\pi r_{+}^3  ~.\label{eeq10}
\end{eqnarray}
In Eq.~(\ref{cg1}), when Rindler parameter goes to zero value then  one obtains the Eq.~(\ref{eeq10}). 
It should be noted that Eq.~(\ref{cg1}) is one of the interesting {results} of this work. 
{The above} Eq.~(\ref{cg1}) represents the thermodynamic volume modified by the non-trivial
Rindler parameter. This means that this geometric term modified the thermodynamic volume of this BH in CG.
This is also quite interesting.

The first law of thermodynamics should read
\begin{eqnarray}
dM &=& T d{\cal S}+ V dP+\chi da ~. \label{eq13}
\end{eqnarray}
where $\chi$ is the physical quantity associated with the parameter $a$. It should 
be defined as
\begin{eqnarray}
\chi &=& \left(\frac{\partial M}{\partial a}\right)_{S,P}
=-\frac{r_{+}^2}{2}\left[1+\frac{3r_{+}\left(a+\frac{8\pi P}{3}r_{+}\right)}
{\sqrt{1-3a^2r_{+}^2-16\pi a Pr_{+}^3}} \right]  ~.\label{eq14}
\end{eqnarray}
{It implies that the Rindler parameter modified the first law of thermodynamics in 
the extended phase space.}

\begin{figure}[h]
  \begin{center}
\subfigure[]{
\includegraphics[width=2.5in,angle=0]{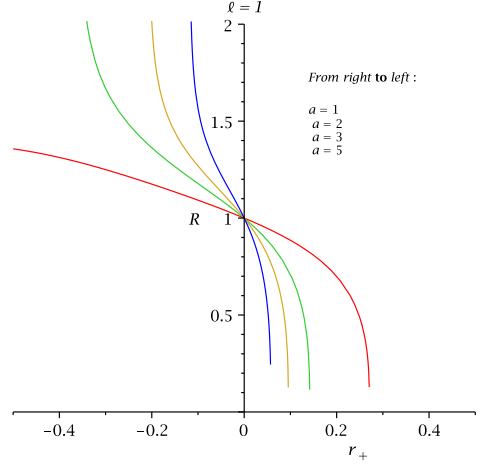}} 
\caption{\label{rrf}\textit{Variation  of ${\cal R}$  with $r_{+}$ for CG  BH for
 different values of Rindler parameter. }}
\end{center}
\end{figure}

Another novel feature of the thermodynamic volume is so called \emph{Reverse Isoperimetric Inequality}~\cite{cvetic11} 
which is satisfied for all BHs except super-entropic BHs~\cite{mann1}. It has been conjectured that the thermodynamic 
volume $V$ and the horizon area $A$  always {satisfy} the isoentropic ratio {i.e.}
\begin{eqnarray}
 {\cal R} &=& \left(\frac{3V}{4\pi} \right)^{\frac{1}{3}} 
 \left(\frac{4\pi}{{\cal A}} \right)^{\frac{1}{2}} \geq 1 ~,\label{rr}
\end{eqnarray}
{For} Schwarzschild-AdS BH, it {is} maximized. But interestingly, in our case this 
ratio is calculated to be 
\begin{eqnarray}
 {\cal R} &=& \left[1-\frac{3ar_{+}}{\sqrt{1-3a^2r_{+}^2-16\pi Par_{+}^3}}\right]^{\frac{1}{3}} \leq 1 ~, \label{rr1}
\end{eqnarray}
and found {that it always} \emph{violates the Reverse Isoperimetric Inequality}. 
{Probably this is a second counter example (after Ref. ~\cite{mann1}) of  
violation of the conjecture ${\cal R} \geq 1$.} It could be seen from  Fig.~\ref{rrf}. 
{From the Eq.~(\ref{rr1}), we could say that the Rindler parameter modifies the 
value of Reverse  Isoperimetric Inequality.} 
It should be noted that when $a=0$, one obtains the {result for} `maximally entropic' 
Schwarzschild-AdS BH~\cite{mann1}.

\begin{figure}[h]
  \begin{center}
\subfigure[]{
\includegraphics[width=2.5in,angle=0]{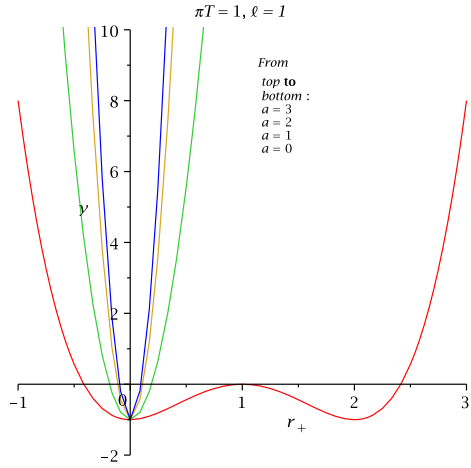}} 
\caption{\label{hpf}\textit{Variation  of 
 $y=r_{+}^4-4 \pi T \ell^2 r_{+}^3+\left(4\pi^2T^2+12 \pi a T+12 a^2\right)\ell^4 r_{+}^2-\ell^4$  
 with $r_{+}$ for CG  BH. It {could be seen} from the figure due to the Rindler 
 {acceleration the} Hawking-Page~(HP) phase transition for CG BH gets modified.}}
\end{center}
\end{figure}

Finally, the Gibbs free energy in the extended phase space should read
\begin{eqnarray}
 G &=& H-TS=\frac{r_{+}}{4}\left[\sqrt{1-3a^2r_{+}^2-16\pi P ar_{+}^3}-3ar_{+}-\frac{8\pi P}{3}r_{+}^2\right]
 ~. \label{cgg1}
\end{eqnarray}
The Gibbs free energy and BH temperature both {depend} on the parameter $a$. The global stability
properties of small and large BHs could be determined by studying the features of $G$. 
From Eq. (\ref{cgg1}) it follows that when $G=0$ {and} $r_{+}=0$,  
the BH is in a pure radiation phase.  The minimum value of the $G$ is at the origin 
which suggests that
\begin{eqnarray}
r_{+}^4-4 \pi T \ell^2 r_{+}^3+\left(4\pi^2T^2+12 \pi a T+12 a^2\right)\ell^4 r_{+}^2-\ell^4 &=& 0  
~.\label{gg1}
\end{eqnarray}
To find the exact numerical value of $r_+$ from the above equation it is a very difficult task. 
Instead, one {could} observe the variation of this function with $r_{+}$ for 
different values of $a$ and $T$ {as presented graphically}~(See the Fig.~\ref{hpf}).
{While} for $a=0$, one {could} easily {determine} 
the value of $r_{+}$ and HP phase transition temperature. 
From Eq.~(\ref{gg1}), we get 
\begin{eqnarray}
r_{+}^4-4 \pi T \ell^2 r_{+}^3+ 4 \pi^2 T^2 \ell^4 r_{+}^2-\ell^4 &=& 0  
~.\label{gg2}
\end{eqnarray}
which is reduced to more simplified form {as}
\begin{eqnarray}
\left( r_{+}^2 -2\pi T \ell^2 r_{+}+\ell^2 \right) \left( r_{+}^2-2 \pi T \ell^2 r_{+}-\ell^2 \right) &=& 0  
~.\label{gg3}
\end{eqnarray}
we discard the second one and from {the} first one we find $r_{+}=\ell=\sqrt{\frac{3}{8 \pi P}}$
when $T=T_{HP}=\frac{1}{\pi \ell}=\sqrt{\frac{8P}{3\pi}}$. 

\begin{figure}[h]
\begin{center}
 \subfigure[ ]{
 \includegraphics[width=2.1in,angle=0]{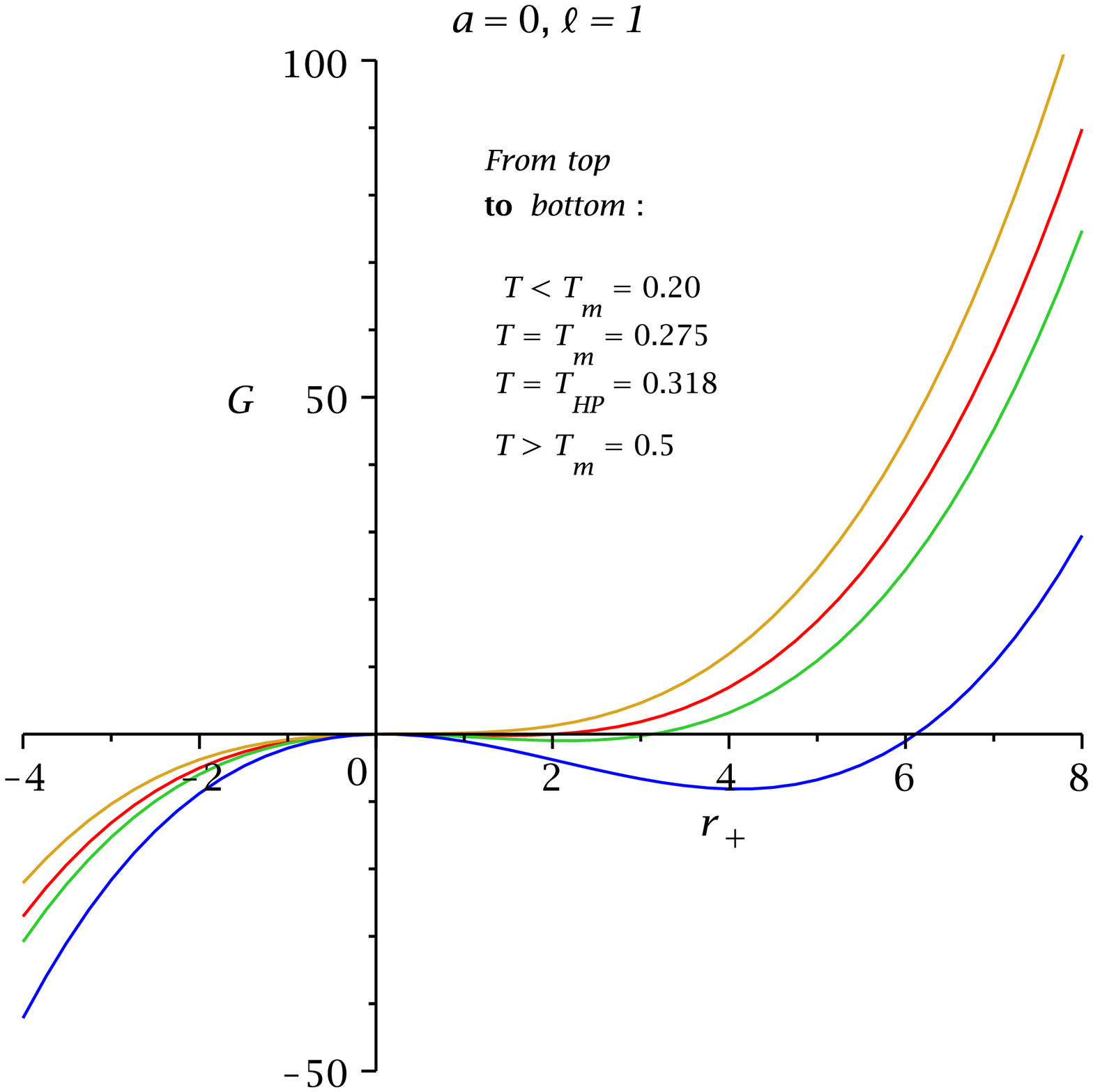}}
 \subfigure[]{
 \includegraphics[width=2.1in,angle=0]{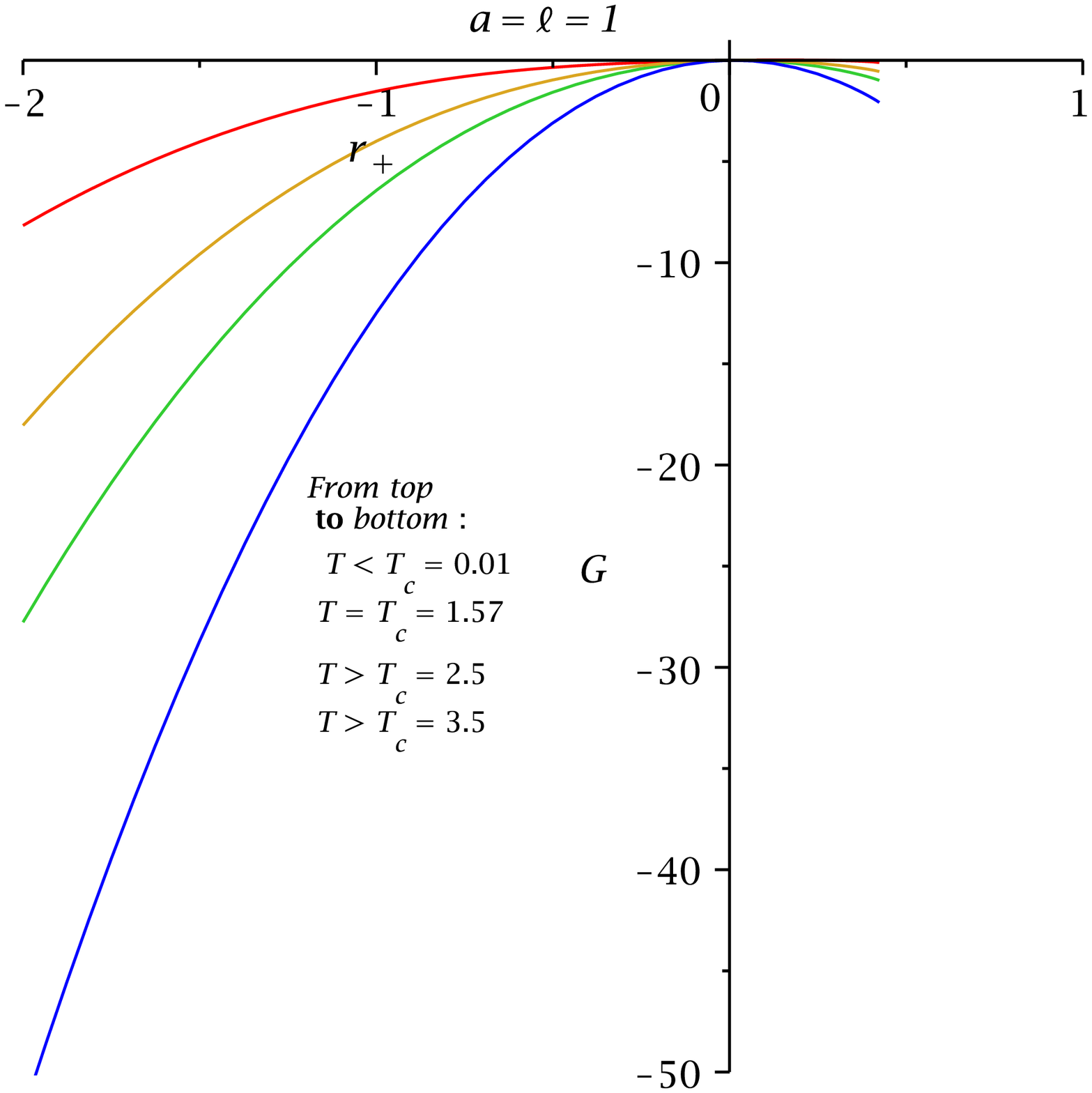}}
 \subfigure[]{
 \includegraphics[width=2.5in,angle=0]{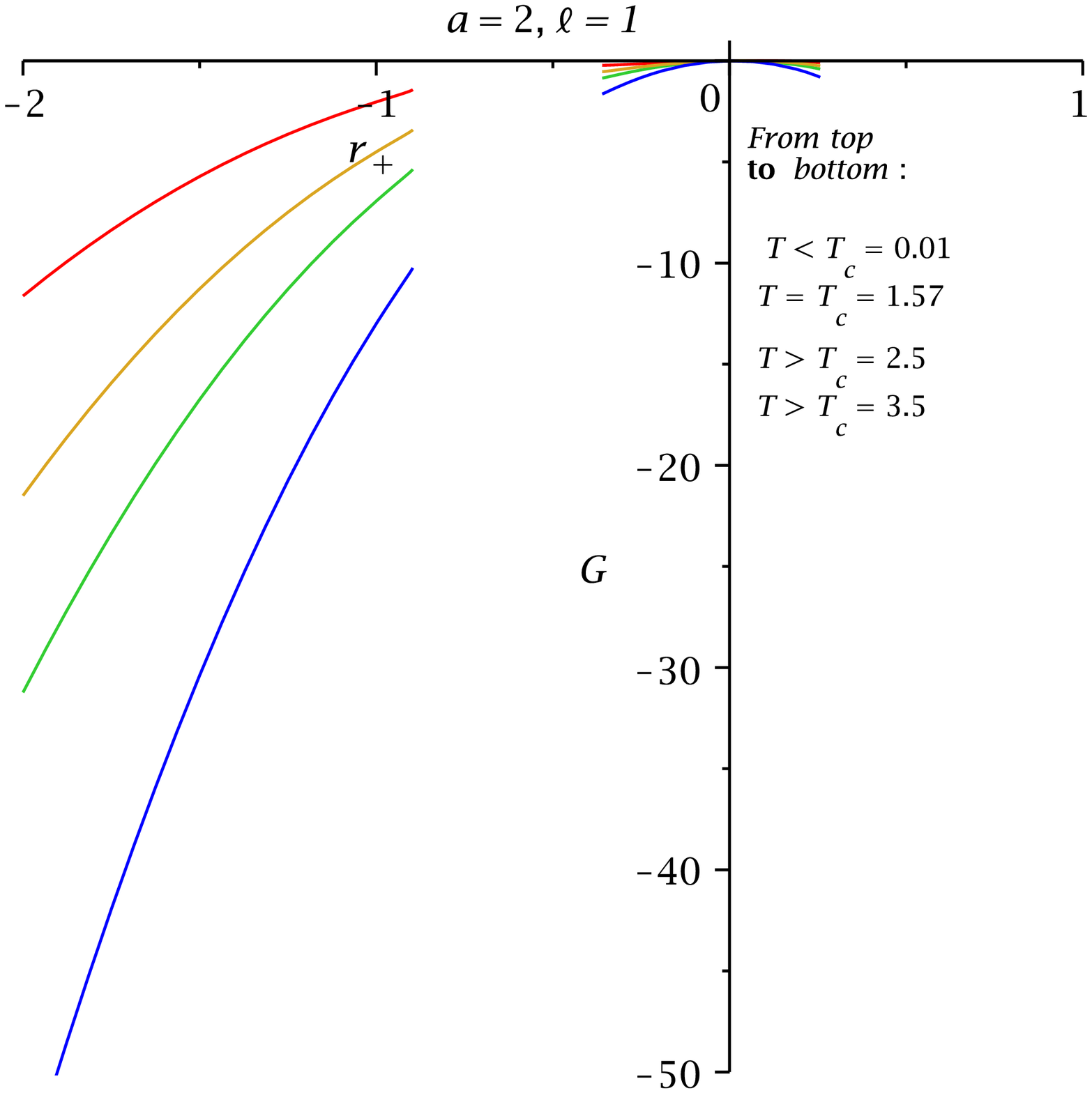}}
 \caption{\label{gf}\textit{Variation  of $G$  with $r_{+}$ for CG BH. It follows from the 
 figure the variation of $G$  with $r_{+}$ without {the parameter}, $a$ and 
 with {the parameter}, $a$ is qualitatively different.}}
\end{center}
\end{figure}

\begin{figure}[h]
  \begin{center}
  \subfigure[]{
\includegraphics[width=2.1in,angle=0]{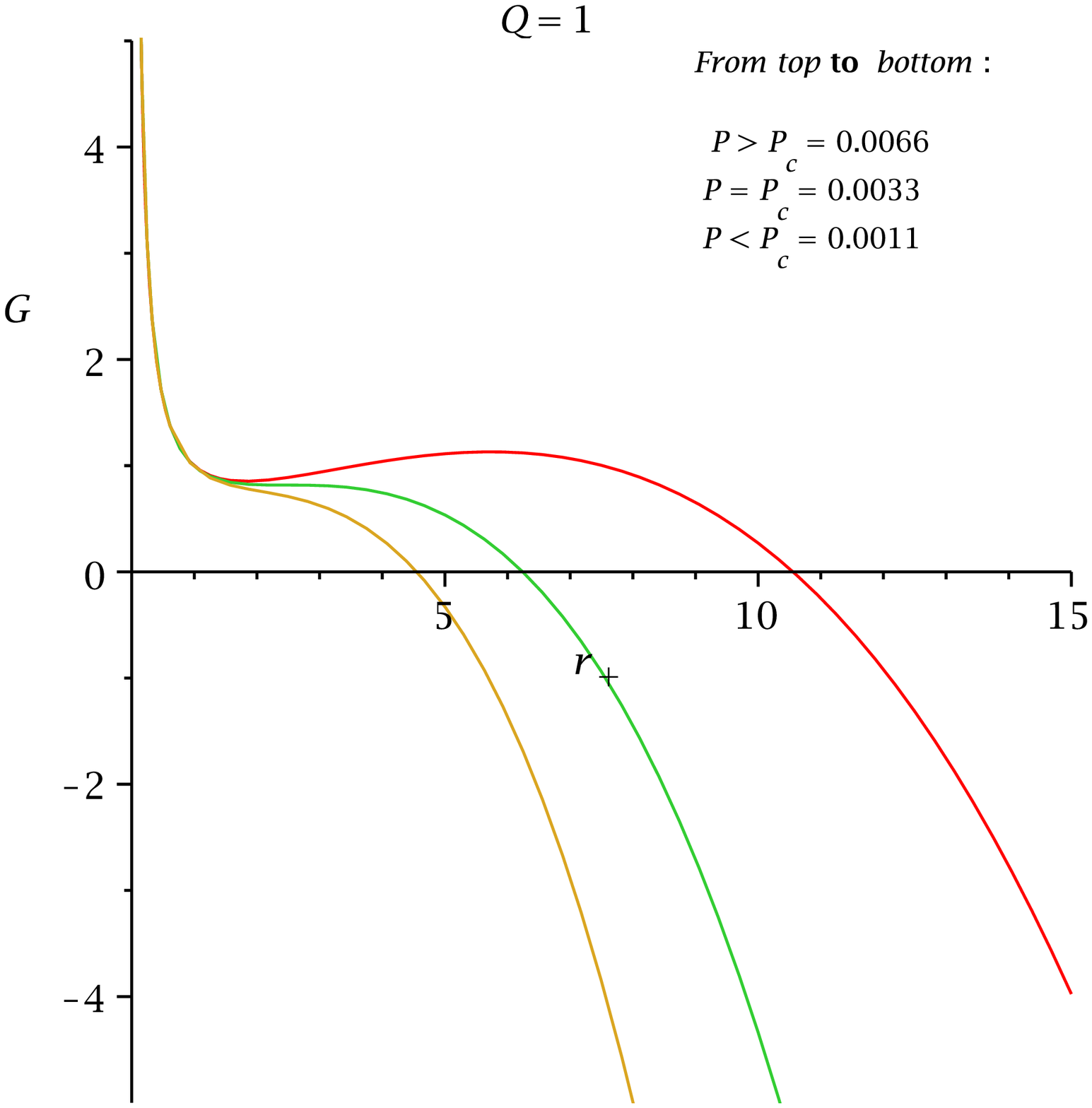}} 
 \subfigure[]{
 \includegraphics[width=2.1in,angle=0]{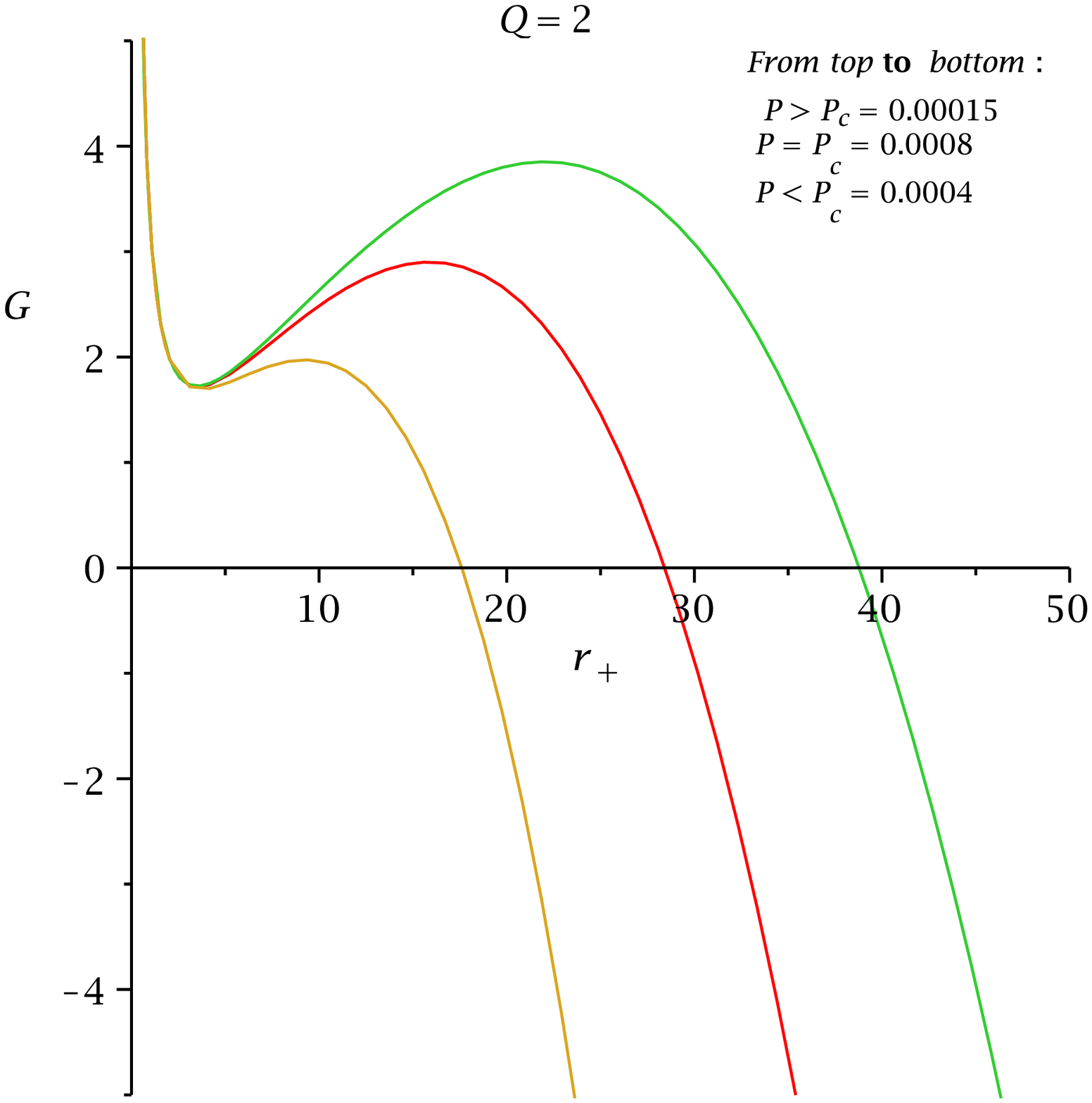}}
 \caption{\label{gf1}\textit{Variation  of $G$  with $r_{+}$ for RN-AdS BH.}}
\end{center}
\end{figure}

\begin{figure}[h]
  \begin{center}
\subfigure[]{
\includegraphics[width=2.1in,angle=0]{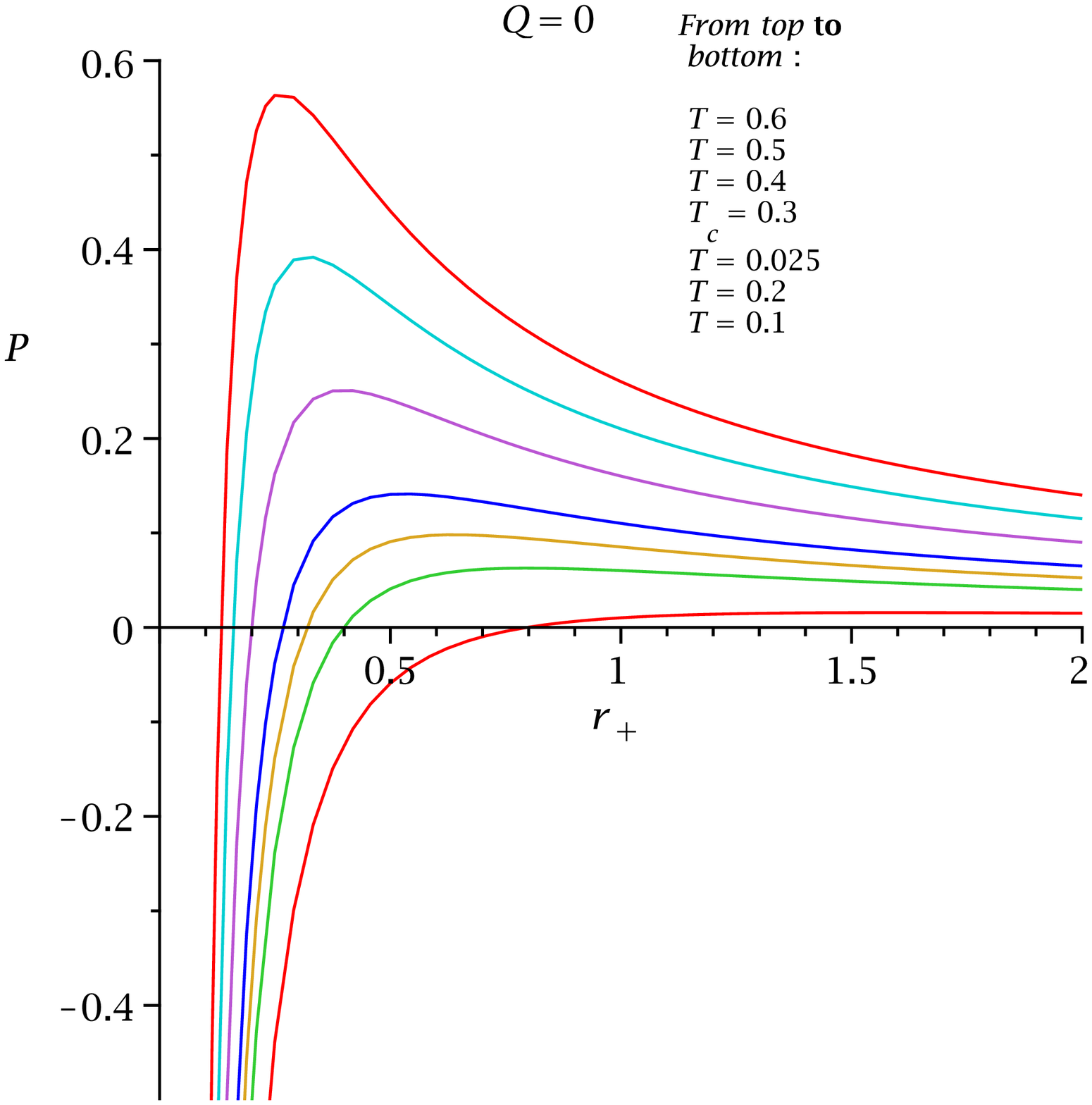}} 
 \subfigure[]{
 \includegraphics[width=2.1in,angle=0]{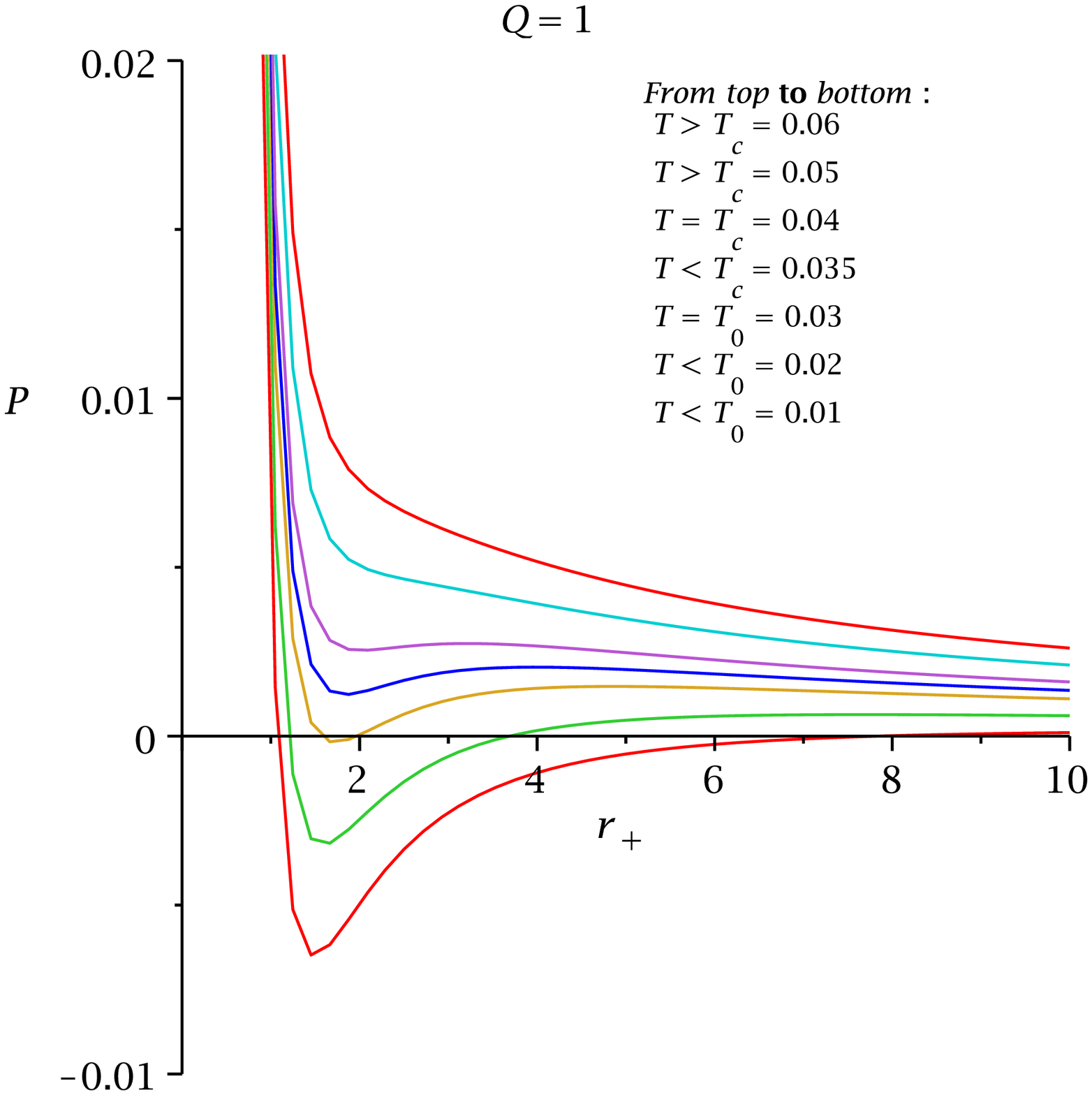}}
 \subfigure[ ]{
 \includegraphics[width=2.1in,angle=0]{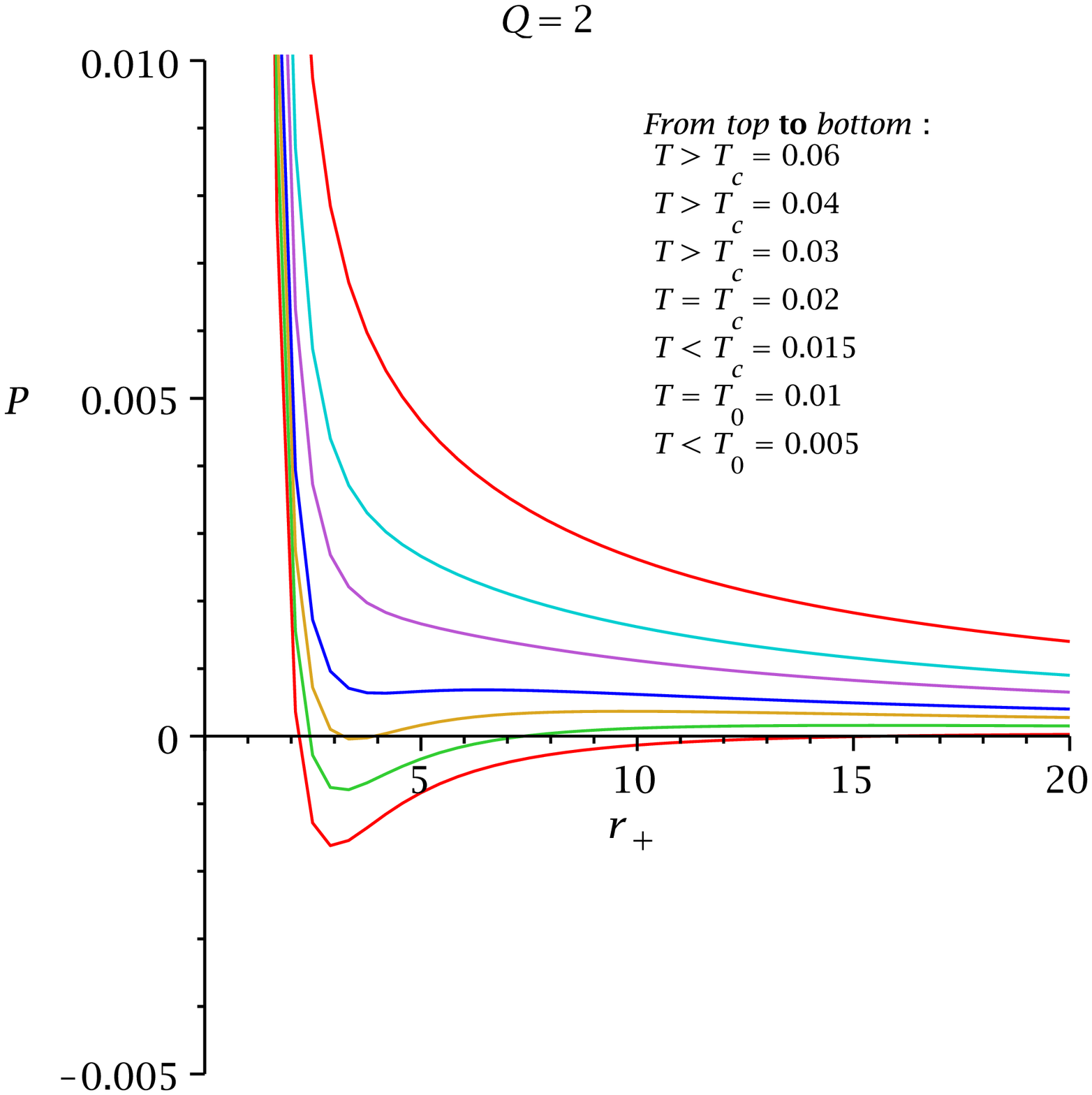}}
 \subfigure[]{
 \includegraphics[width=2.1in,angle=0]{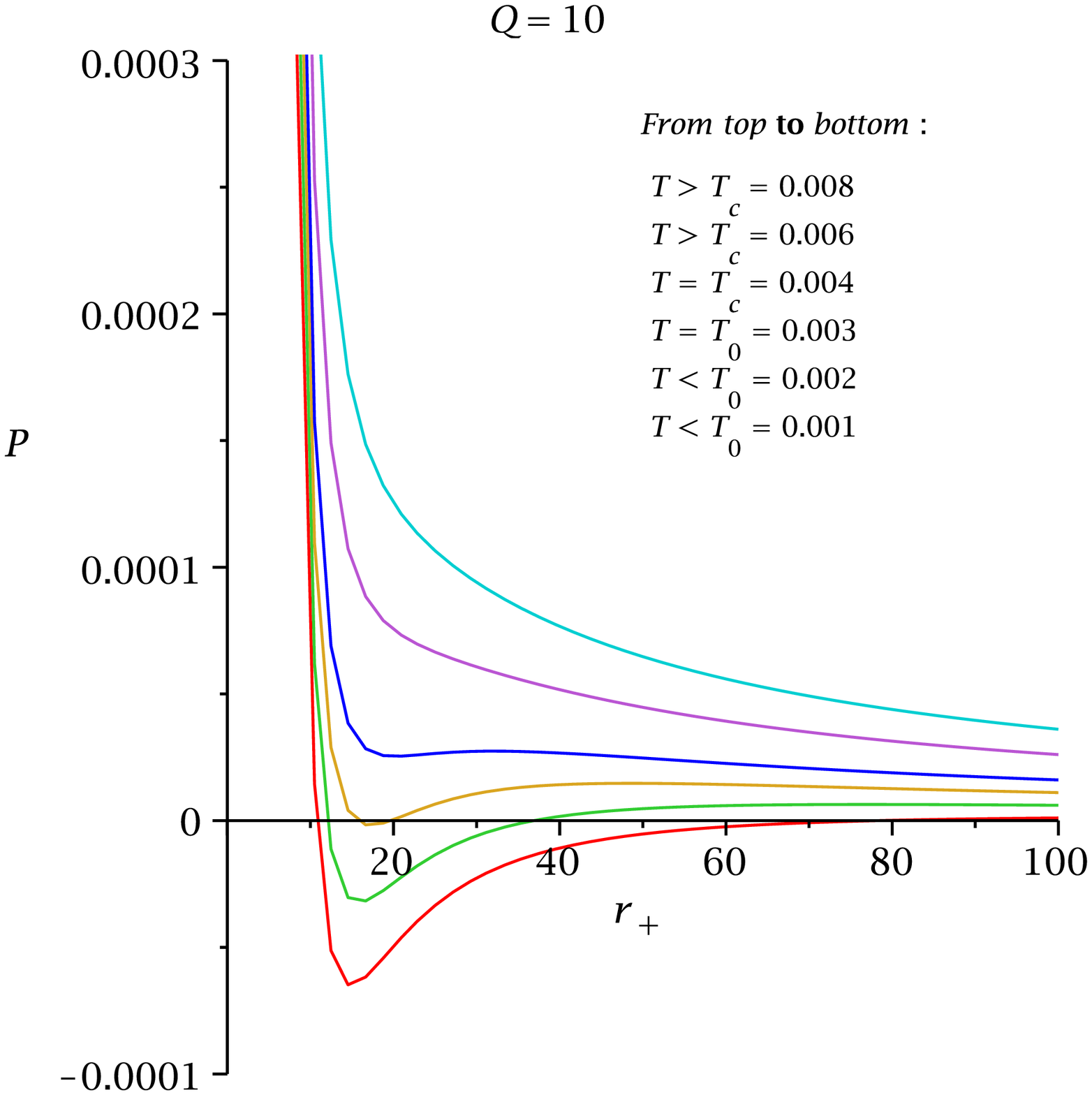}}
 \caption{\label{fg}\textit{Variation  of $P$  with $r_{+}$ for Schwarzschild-AdS BH and RN-AdS space-time. 
 {The above figures indicate that the $P-r_{+}$ diagram for $Q=0$ and $Q=1,2,3$ are qualitatively 
 different in nature. For $Q=0$, the isotherms have no inflection point. Whereas for $Q=1$, $Q=2$ and $Q=3$, the 
 isotherms have inflection point. This is the main differences between Schwarzschild-AdS BH and RN-AdS space-time. } }}
\end{center}
\end{figure}

Where $T_{HP}$ is called the famous Hawking-Page~(HP) critical phase transition temperature~\cite{haw83}. 
For $T>T_{HP}$, the large BH is globally stable and for $T<T_{HP}$, the small BH is  
thermodynamically unstable, while the larger one is locally 
stable~\cite{euro}. But in our case, due to the Rindler parameter we expect that the $T_{HP}$ {gets}
modified by the factor $a$. We {could} not find the exact HP temperature due to the quartic nature of 
the {polynomial} equation, but we {could} say
that it must be a function of Rindler parameter. In Fig.~\ref{gf}, we have drawn the Gibbs free energy 
for different values of temperature for CG BH.
In Fig.~\ref{gf1}, we have plotted the Gibbs free energy for different values of
temperature for RN BH in comparison with CG BH.
Now we turn to the main work. Using Eqs.~(\ref{eq3}) and (\ref{eq8}), the Hawking temperature {could} 
be rewritten as 
\begin{eqnarray}
T &=& \frac{1}{4\pi r_{+}} \left[a r_{+}+8\pi P r_{+}^2+\sqrt{1-3a^2r_{+}^2-16\pi a Pr_{+}^3}\right]
~. \label{eq3.1}
\end{eqnarray}
Using this equation one {could} obtain the BH equation of state as
\begin{eqnarray}
64 \pi^2 r_{+}^4P^2+32\pi r_{+}^3\left(a-2\pi T\right)P+\left(16\pi^2r_{+}^2T^2-8\pi ar_{+}^2T+4a^2r_{+}^2-1\right) &=& 0  
~.\label{cc1}
\end{eqnarray}
This is a quadratic equation of $P$. Solving this equation one {could} find the 
\emph{equation of state} for this AdS BH 
\begin{eqnarray}
P &=& \frac{T}{2r_{+}}-\frac{a}{4\pi r_{+}} \pm \frac{\sqrt{1-8\pi a T r_{+}^2}}{8\pi r_{+}^2}  
~.\label{cg2}
\end{eqnarray}
The variation of $P-V$ diagram {could} be seen from the Fig.~\ref{fg} for Schwarzschild-AdS BH.
{For} RN-AdS BH \&  CG BH, {it could be seen} from Fig.~\ref{fg1} \& Fig.~\ref{fgg1}. 
 One {could}  easily {observe}  from this plot {that}  
due to the presence of the Rindler parameter there has been a deformation of shape  of the isotherms in the 
$P-V$ diagram in comparison with charged-AdS BH and chargeless-AdS BH. 

It follows from the Eq.~(\ref{cg2}) {that}  due to the presence of the Rindler term  the 
\emph{BH equation of state} { gets} modified. 
{First,} we consider the lower sign~(the negative one) for $P-V$ criticality. 
The $P-V$ criticality for upper sign should be considered in the Appendix {section}.
In terms of specific volume $v=2r_{+}$~\footnote{where $r_{+}$ is the root of the equation 
$16\pi aPr_{+}^9+12a^2r_{+}^8 -(1+24aPV)r_{+}^6-\left(\frac{9V}{2\pi}\right)a^2r_{+}^5+\left[\left(\frac{3V}{2\pi}\right)
+\frac{9PV^2}{\pi}a\right]r_{+}^3+\left(\frac{27a^2V^2}{16\pi^2}\right)r_{+}^2-\left(\frac{3V}{4\pi}\right)^2=0$, In 
the limit $a=0$, $r_{+}=\left(\frac{3V}{4\pi}\right)^{\frac{1}{3}}$.}, one  obtains the {BH} equation of 
state {as}
\begin{eqnarray}
P &=& \frac{T}{v}-\frac{a}{2\pi v}-\frac{\sqrt{1-2\pi av^2T}}{2\pi v^2}  ~.\label{cg3}
\end{eqnarray}
In the limit $a=0$, one has the equation of state for Schwarzschild-AdS BH~\cite{nata} as
\begin{eqnarray}
P &=& \frac{T}{v}-\frac{1}{2\pi v^2}  ~, \label{cg3.1}
\end{eqnarray}
where $v=2r_{+}=2\left(\frac{3V}{4\pi}\right)^{\frac{1}{3}}$.

\begin{figure}[h]
  \begin{center}
\subfigure[]{
\includegraphics[width=2.1in,angle=0]{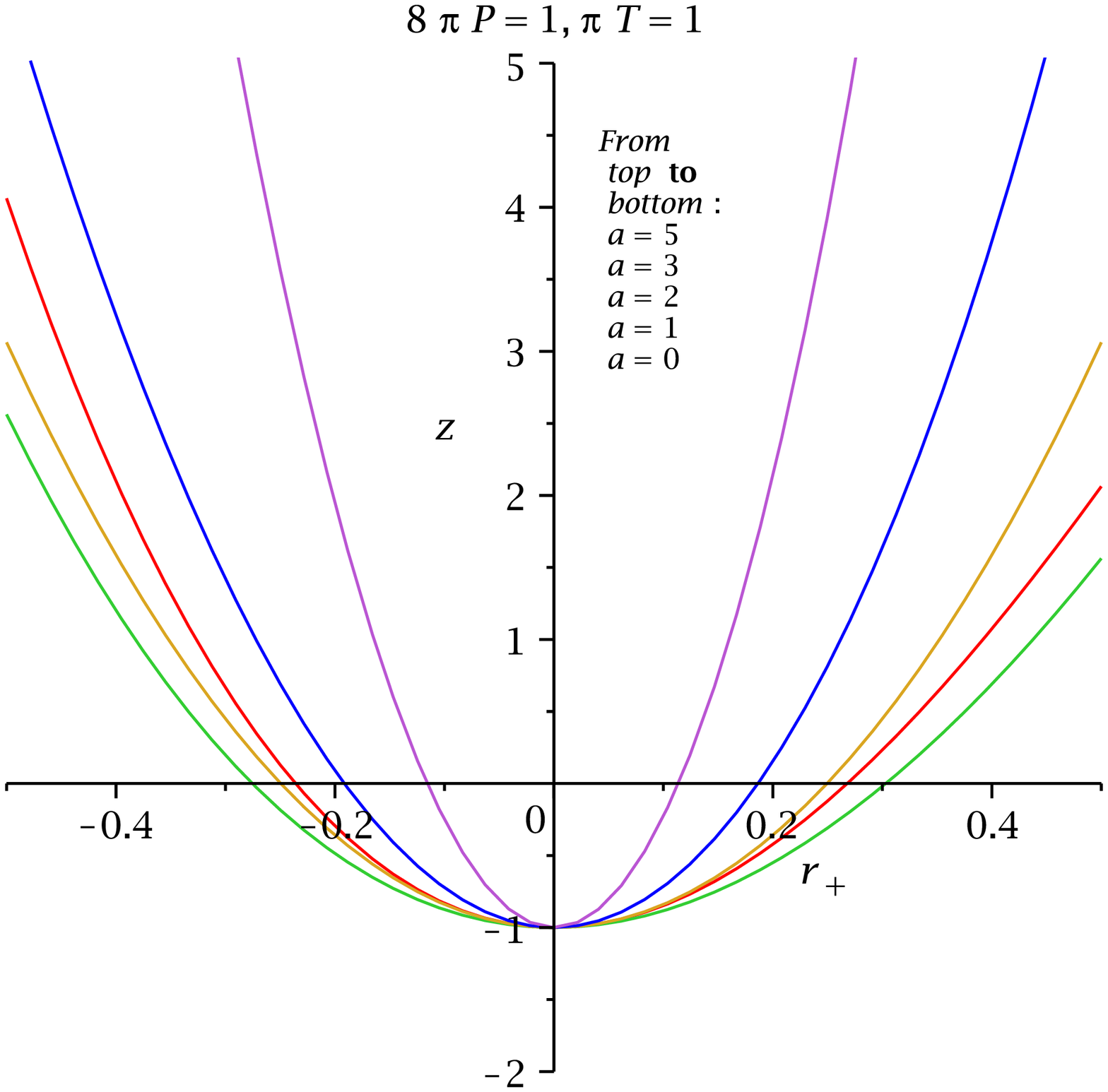}} 
 \subfigure[]{
 \includegraphics[width=2.1in,angle=0]{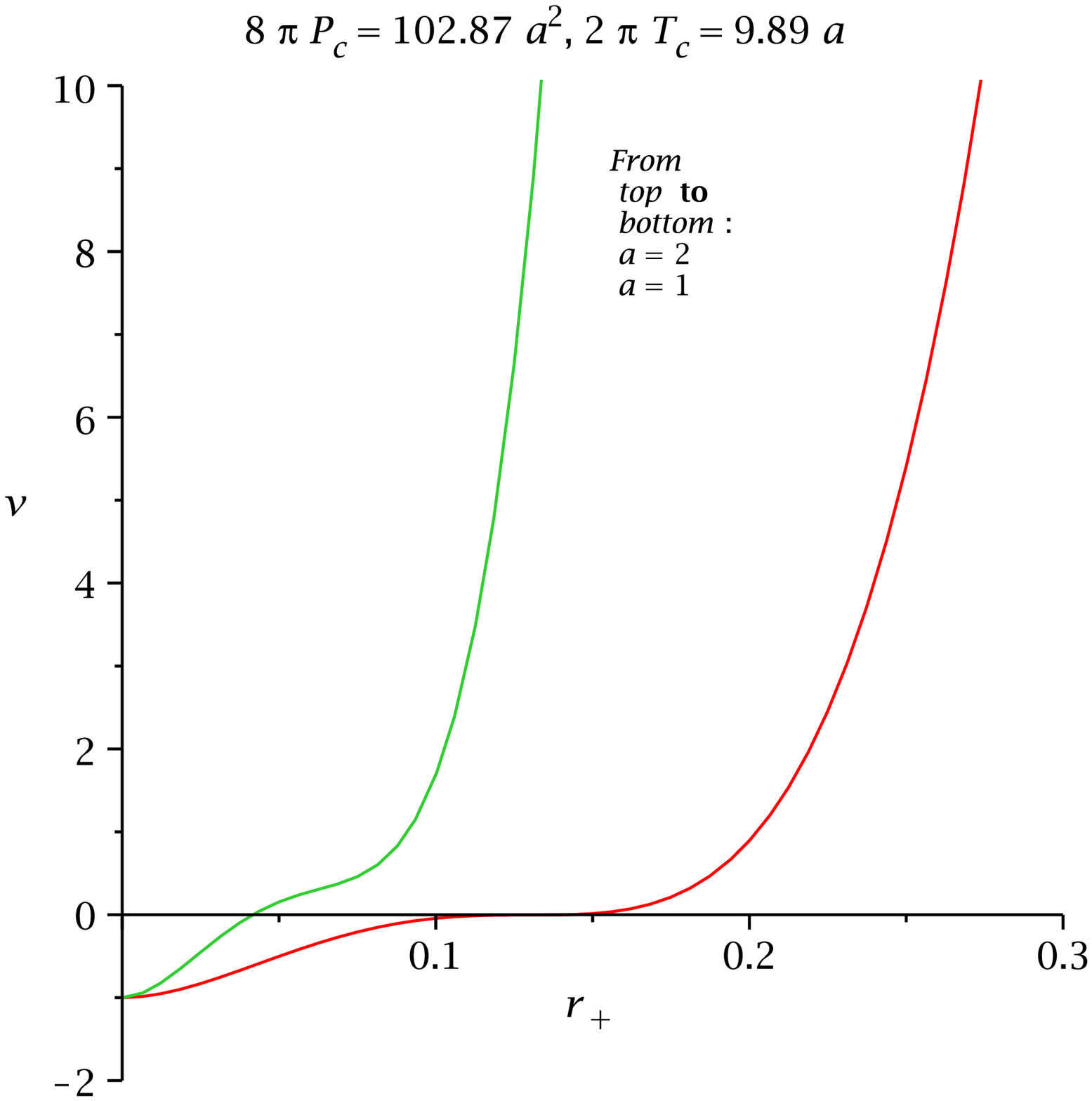}}
 \caption{\label{lsf}\textit{Variation  of 
 $z= (8\pi P)^2 r_{+}^4+32\pi P\left(a-2\pi T\right)r_{+}^3+\left(16\pi^2T^2-8\pi aT+4a^2\right)r_{+}^2-1$ and
 $v= (8\pi P_{c})^2 r_{+}^4+32\pi P\left(a-2\pi T_{c}\right)r_{+}^3+\left(16\pi^2T_{c}^2-8\pi aT_{c}+4a^2\right)r_{+}^2-1$
 with $r_{+}$ for CG  BH.}}
\end{center}
\end{figure}

\begin{figure}[h]
  \begin{center}
\subfigure[]{
\includegraphics[width=2.1in,angle=0]{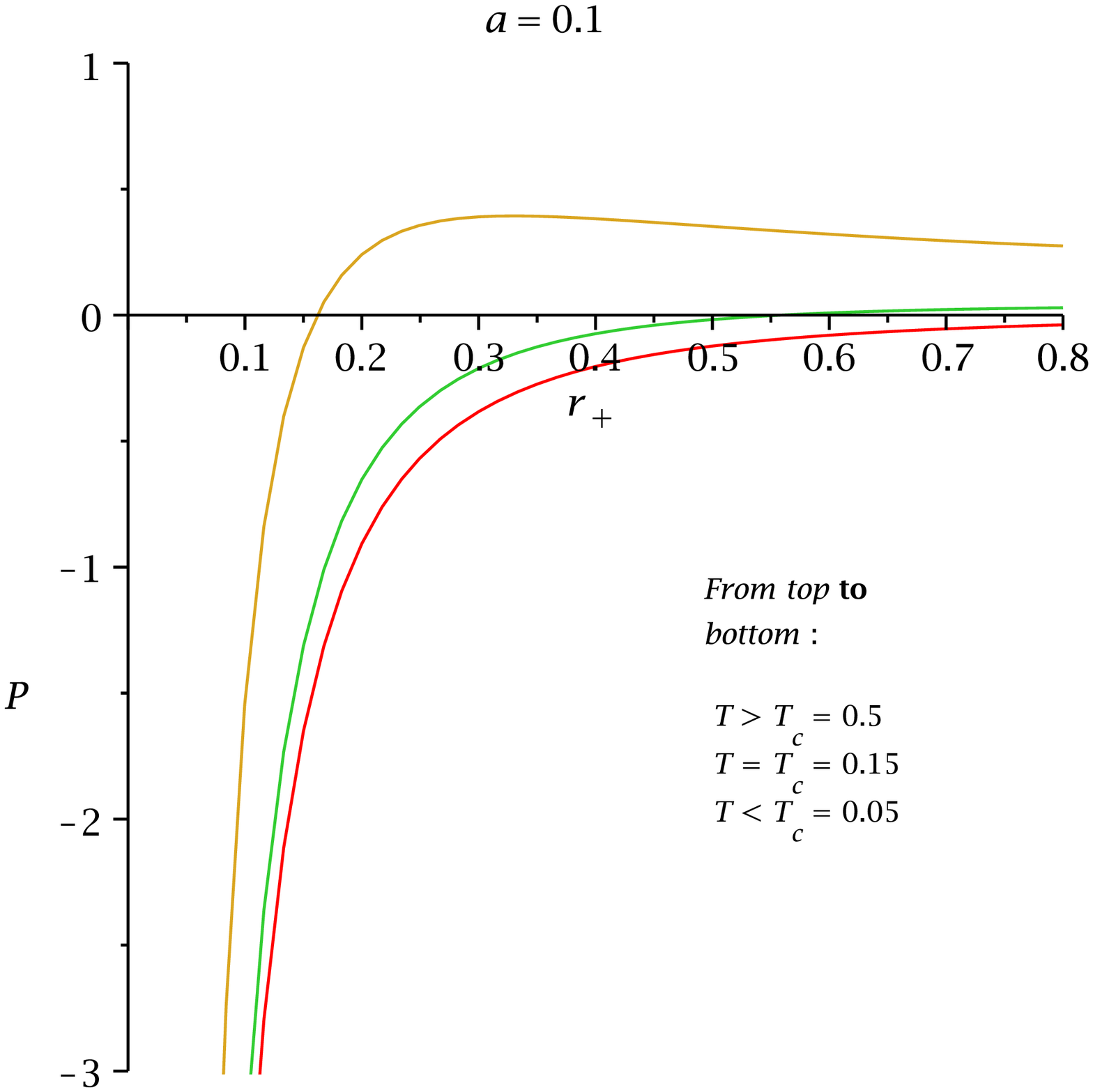}} 
\subfigure[]{
 \includegraphics[width=2.1in,angle=0]{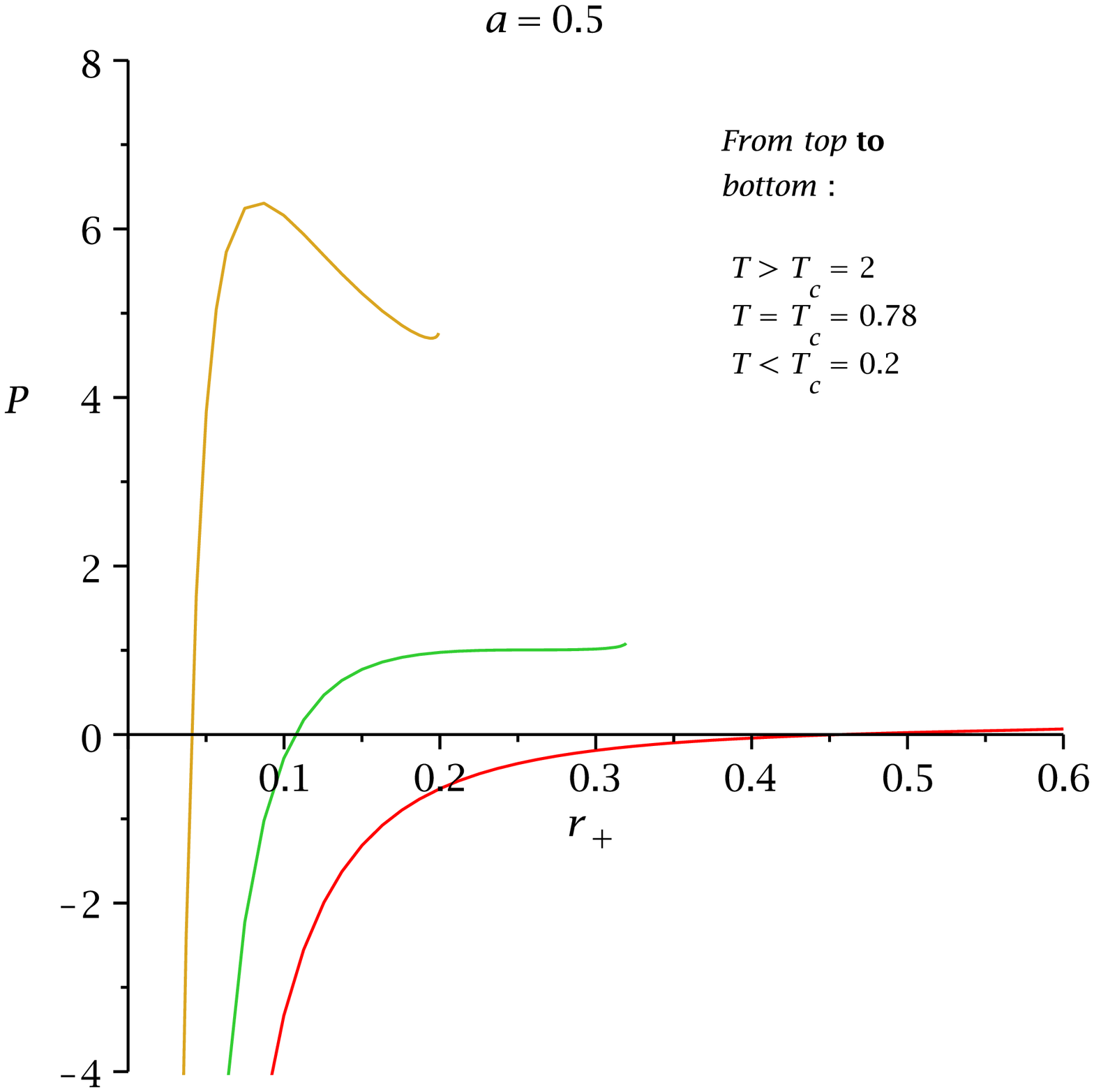}}
\subfigure[]{
 \includegraphics[width=2.1in,angle=0]{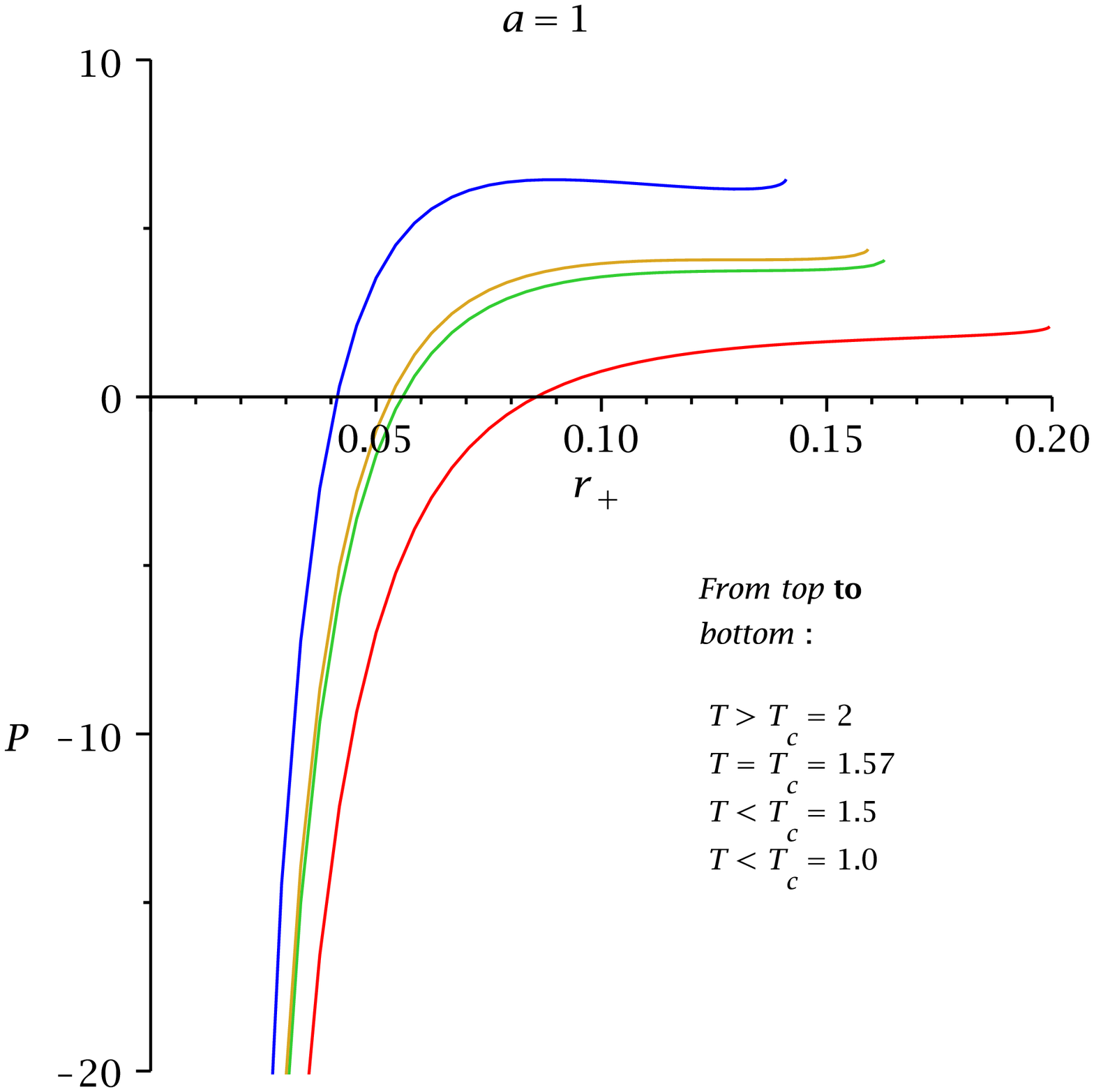}}
 \subfigure[]{
 \includegraphics[width=2.1in,angle=0]{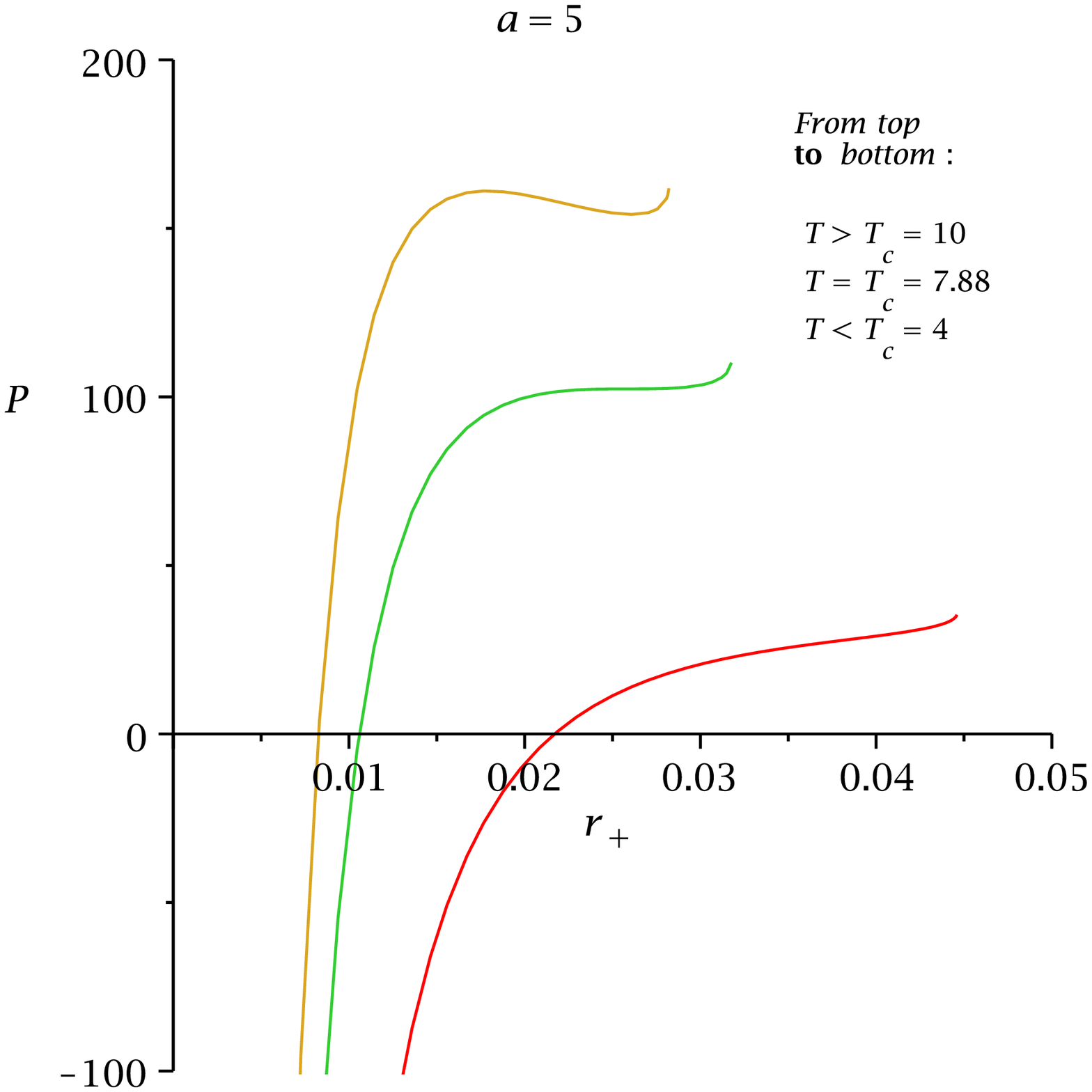}}
\caption{\label{fg1}\textit{Variation  of $P$  with $r_{+}$ for various values of $a$. }}
\end{center}
\end{figure}

\begin{figure}[h]
  \begin{center}
 \subfigure[ ]{
 \includegraphics[width=2.1in,angle=0]{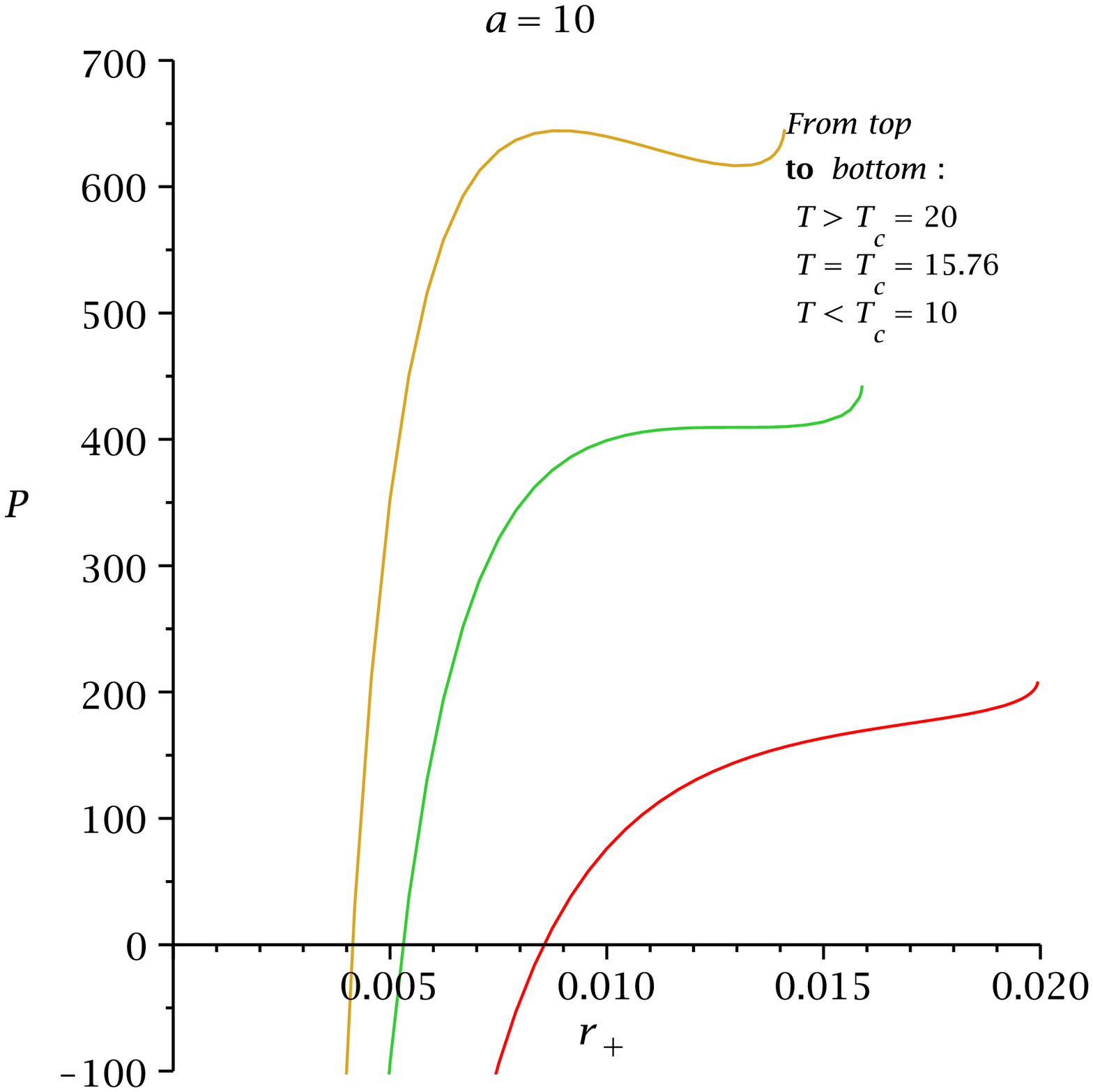}}
 \subfigure[]{
 \includegraphics[width=2.1in,angle=0]{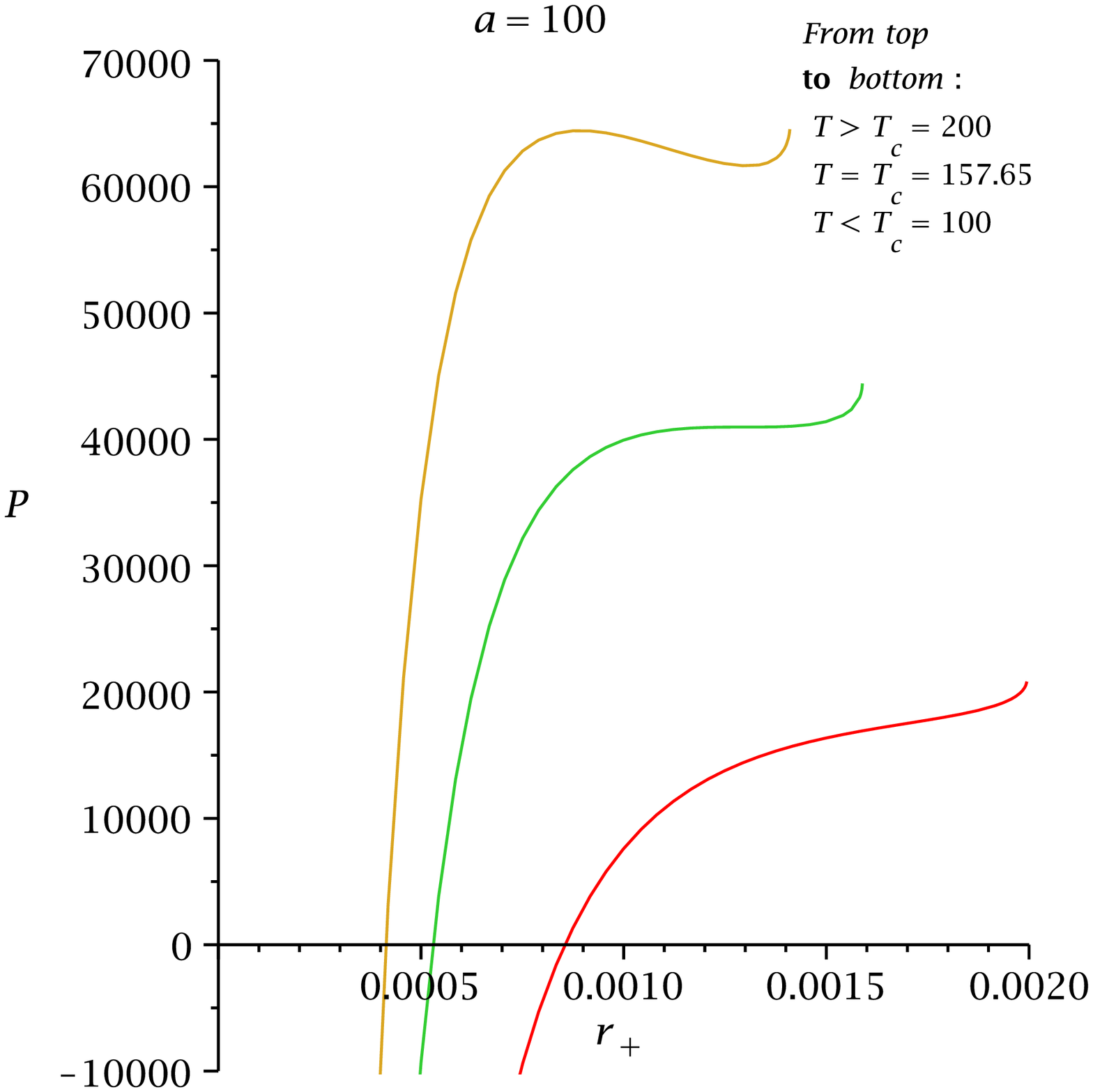}} 
\caption{\label{fgg1}\textit{Variation  of $P$  with $r_{+}$ for various values 
of $a$. }}
\end{center}
\end{figure}

The variation of $P-r_{+}$ diagram {could} be seen from the Fig.~\ref{fg}-$a$. 
{Each isotherm corresponds to a maximum value of $P$} at $r_{+}=\frac{1}{2\pi T}$. 
As we have mentioned earlier, for a particular temperature $T=T_{min}$, 
the value of $r_{+}=r_{min}=\frac{1}{2\pi T_{min}}=\frac{\ell}{\sqrt{3}}=\frac{1}{\sqrt{8\pi P}}$. It should be noted that 
$G$ exhibits an inflection point at $r_{min}$. For greater value of this temperature  {we would find}
the radii {for one }large and  {one} small BH. 
From Eq.~(\ref{cc1}), we can obtain the radii of large and small BH for CG BH. 
{ It is a very difficult task to determine the exact roots of 
the quartic equation numerically. So we can plot this function graphically for 
various values of $a$, $P$ and $T$~(See Fig.~\ref{lsf}) }. 
But in the limit $a=0$, the Eq.~(\ref{cc1}) reduces to the {following} form
\begin{eqnarray}
64 \pi^2 P^2 r_{+}^4-64\pi^2 T P r_{+}^3+16\pi^2T^2r_{+}^2-1 &=& 0  ~.\label{cc2}
\end{eqnarray}
{It could be more simplified  as}
\begin{eqnarray}
\left(8 \pi P r_{+}^2-4\pi T r_{+}+1\right) \left(8 \pi P r_{+}^2-4\pi T r_{+}-1\right) &=& 0  
~.\label{cc3}
\end{eqnarray}
The first equation gives the radii of large and small Schwarzschild-AdS BH which is given by 
\begin{eqnarray}
 r_{large} &=& \frac{T}{4P}\left[1+\sqrt{1-\frac{2P}{\pi T^2}} \right]  \\
 r_{small} &=& \frac{T}{4P}\left[1-\sqrt{1-\frac{2P}{\pi T^2}} \right]~ \label{cc4}
\end{eqnarray}
where $T_{min}=\sqrt{\frac{2P}{\pi}}$. The second one gives another set of radii 
of large and small BH
\begin{eqnarray}
 r_{large} &=& \frac{T}{4P}\left[1+\sqrt{1+\frac{2P}{\pi T^2}} \right]  \\
 r_{small} &=& \frac{T}{4P}\left[1-\sqrt{1+\frac{2P}{\pi T^2}} \right]~ \label{cc5}
\end{eqnarray} 
{This is unphysical because in the discriminant part the value of $T$ becomes 
imaginary}. Since we are {unable} to 
{find} the exact root of $r_{+}$ due to {the} quartic nature 
of Eq.~(\ref{cc2}) {if we set $T=\frac{a}{2\pi}$, and assuming if it is 
the HP phase transition temperature for CG BH,  the Eq.~(\ref{cc1}) reduces} 
to the following form
\begin{eqnarray}
64 \pi^2 P^2 r_{+}^4+4a^2r_{+}^2-1 &=& 0  ~\label{cc6}
\end{eqnarray}
Using {the above equation}, one obtains the radii of large and small BH for CG gravity
\begin{eqnarray}
 r_{large} &=& \sqrt{\left|\frac{\sqrt{a^4+16\pi^2P^2}+a^2}{32\pi^2P^2}\right|}  \\
 r_{small} &=& \sqrt{\left|\frac{\sqrt{a^4+16\pi^2P^2}-a^2}{32\pi^2P^2}\right|}  ~ \label{cc7}
\end{eqnarray}
{Hence one may } conclude that the HP {phase transition} temperature for 
CG BH must be the {function of Rindler parameter i.e.} $T_{HP}=f(a)$. 
{This further} implies that the HP {phase transition}  temperature 
depends on the Rindler parameter. The discussion of fluid {analogue} of Schwarzschild-AdS BH 
{could} be found in Ref.~\cite{nata}.  

The critical constants could be determined by applying the following conditions at the 
inflection point
\begin{eqnarray}
\frac{\partial P}{\partial v}|_{T=T_{c}} &=& 0  ~. \label{cg4}
\end{eqnarray}
\begin{eqnarray}
\frac{\partial^2 P}{\partial v^2}|_{T=T_{c}} &=& 0  ~. \label{cg5}
\end{eqnarray}
Solving Eq.~(\ref{cg4}), one can obtain
\begin{eqnarray}
T_{c} &=& \frac{a}{2\pi}+ \frac{1-\pi av_{c}^2T_{c}}{\pi v_{c}\sqrt{1-2\pi av_{c}^2T_{c}}}  
~. \label{cg6}
\end{eqnarray}
{and} solving Eq.~(\ref{cg5}), one can find
\begin{eqnarray}
T_{c} &=& \frac{a}{2\pi}+ \frac{3-9\pi av_{c}^2T_{c}+4\pi^2a^2v_{c}^4T_{c}^2}
{2\pi v_{c}\left(1-2\pi av_{c}^2T_{c}\right)^{3/2}}  
~. \label{cg7}
\end{eqnarray}
In the limit $a=0$, one obtains the critical constants for Schwarzschild-AdS BH
\begin{eqnarray}
P_{c} &=& \frac{1}{2\pi v_{c}^2} \\
T_{c} &=& \frac{1}{\pi v_{c}} ~. 
\label{cg7.1}
\end{eqnarray}
If we choose $v_{c}=1$, we find {that} the critical values are $P_{c} = \frac{1}{2\pi}$ and 
$T_{c} = \frac{1}{\pi}$. The interesting case {that we observe an HP phase transition between
small and large BHs~\cite{haw83}.}  Witten~\cite{witten} explained {that} the HP phase 
transition is  dual to the QCD confinement/deconfinement phase transition.

Using Eq.~(\ref{cg6}) and Eq.~(\ref{cg7}), one {could derive} the critical Hawking temperature
\begin{eqnarray}
T_{c} &=& \frac{1}{3\pi av_{c}^2} ~. \label{cg8}
\end{eqnarray}
Using Eq.~(\ref{cg6}) and Eq.~(\ref{cg8}), one obtains the critical volume
\begin{eqnarray}
v_{c} &=& \frac{3\sqrt{2}-2\sqrt{3}}{3a} ~. \label{cg9}
\end{eqnarray}
Finally, using Eq.~(\ref{cg3}) and Eq.~(\ref{cg8}), we {could} find the 
critical pressure
\begin{eqnarray}
P_{c} &=& \frac{\sqrt{3}}{2\pi v_{c}^2} ~. \label{cg10}
\end{eqnarray}
In terms of $a$, the critical values are 
\begin{eqnarray}
P_{c} &=& \frac{3\sqrt{3}}{4\pi \left(5-2\sqrt{6}\right)}a^{2} \\
v_{c} &=& \frac{3\sqrt{2}-2\sqrt{3}}{3a} \\
T_{c} &=& \frac{a}{2\pi \left(5-2\sqrt{6}\right)}
~. \label{cg11}
\end{eqnarray}
It may be noted that $P_{c}$, $v_{c}$ and $T_{c}$ {are 
strictly dependent} upon the parameter $a$. 

From the critical constants, one {could} derive the critical ratio for CG BH
\begin{eqnarray}
\rho_{c} &=& \frac{P_{c} v_{c}}{T_{c}}= \frac{\sqrt{3}}{2}\left(3\sqrt{2}-2\sqrt{3}\right) ~. \label{cg12}
\end{eqnarray}
which is a constant value as it {was already} expected and which was 
{found} earlier for 
charged-AdS BH~\cite{david12} as
\begin{eqnarray}
\rho_{c} &=& \frac{P_{c} v_{c}}{T_{c}}= \frac{3}{8} ~. \label{cg13}
\end{eqnarray}
For Schwarzschild-AdS BH, the $\rho_c$ is calculated to be  
\begin{eqnarray}
\rho_{c} &=& \frac{P_{c} v_{c}}{T_{c}}= \frac{1}{2} ~. \label{cg14}
\end{eqnarray}
using Eq.~(\ref{cg7.1}).

It should be noted that for Schwarzschild-AdS BH the isotherm in $P-r_{+}$ diagram is quite 
different from charged-AdS BH. It may be observed from the Fig.~(9-a). The only interesting feature 
{ that happens} in the Schwarzschild-AdS spacetime is  the HP
phase transitions between small and large BH {that} we have described earlier. 

Therefore the ratio of $\rho_c$ for these {BHs} should read 
\begin{eqnarray}
\rho_{c}^{CG}:\rho_{c}^{Sch-AdS}:\rho_{c}^{RN-AdS} &=& 0.67:0.50:0.37 ~. \label{cg15}
\end{eqnarray}
It immediately follows that $\rho_{c}^{CG}>\rho_{c}^{Sch-AdS}>\rho_{c}^{RN-AdS}$. 
The ``law of corresponding states'' becomes 
\begin{eqnarray}
 2\Theta &=& \left(3\sqrt{2}-2\sqrt{3}\right)\Phi\left[\sqrt{3}\Xi+\frac{\sqrt{1-\frac{2}{3}\Theta\Phi^2}}{\Phi^2}\right]  
 ~. \label{cg16}
\end{eqnarray}
where $\Theta$, $\Phi$ and $\Xi$ {could} be defined as 
\begin{eqnarray}
\Theta &=& \frac{T}{T_{c}}\\
\Phi &=& \frac{v}{v_{c}}\\
\Xi &=&  \frac{P}{P_{c}} ~. \label{cg17}
\end{eqnarray}
and these quantities like $\Theta$, $\Phi$ and $\Xi$ are called the \emph{reduced temperature}, the \emph{reduced volume} and 
the \emph{reduced pressure} respectively. Thus the Eq.~(\ref{cg16}) is called the \emph{reduced equation of state}.

\section{\label{dis} Conclusions}
{We have analyzed the critical behaviour of CG BH in four dimensions in the context of extended 
phase space, a thermodynamic model which has garnered increasing interest over the past few years for providing 
more complete description of the thermodynamic processes of BHs in spacetimes with cosmological constants.
The spacetime geometry that we have considered in this work contained a non-trivial Rindler term which 
produces anomolous acceleration in a geodesics of a test particle. It could be  observed in various 
anomolous systems like star-galaxy, Earth-Satellite etc.}
 
{We have examined the effect of this novel parameter on the thermodynamic behaviour}. 
Due to this novel parameter, we observed that in the $P-V$ diagram, the {silhouette of the isotherms} of 
CG BH is quite distinguished from the charged-AdS BH 
and Schwarzschild-AdS BH. We also derived the {BH thermodynamic equation of state, 
critical constant,  Reverse Isoperimetric Inequality, first law of thermodynamics, Hawking-Page phase transition 
and Gibbs free energy}. {We speculated that all these thermodynamic quantities are strictly 
dependent upon this novel parameter. More importantly, the Rindler acceleration term modified the first law of BH 
thermodynamics. We also observed that the HP phase transition temperature is a function of the said parameter. }  
Furthermore, we computed the critical ratio for CG BH in comparison with RN-AdS BH and it 
{complied with} the inequality $\rho_{c}^{CG}>\rho_{c}^{Sch-AdS}>\rho_{c}^{RN-AdS}$. 
{Interestingly, the critical ratio is independent of the Rindler acceleration.} 
Furthermore, we derived  the reduced equation of state in terms of the reduced temperature, the 
reduced volume and the reduced pressure respectively. 
{To summarize, the Rindler acceleration has an effect on the thermodynamic properties of
the CG in $d=4$.}

\section{Appendix}


In the appendix section,  we have examined {that} the $P-V$ criticality for the 
BH equation state of CG BH {which} corresponds to
\begin{eqnarray}
P &=& \frac{T}{2r_{+}}-\frac{a}{4\pi r_{+}} + \frac{\sqrt{1-8\pi a T r_{+}^2}}{8\pi r_{+}^2}  
~.\label{ax1}
\end{eqnarray}
{In} terms of specific volume $v=2r_{+}$, the equation of state {could} 
be written as
\begin{eqnarray}
P &=& \frac{T}{v}-\frac{a}{2\pi v}+\frac{\sqrt{1-2\pi av^2T}}{2\pi v^2}  ~.\label{ax2}
\end{eqnarray}
It should be noted that when $a=0$, we {could} not find the BH equation of state 
for Schwarzschild-AdS BH. Yet, we {would} see what happens in the critical constants
if we use the {positive} sign instead of {negative} sign. 
Doing all the calculations as we have done previously, we {have  found that} the 
critical constants as 
\begin{eqnarray}
P_{c} &=& -\frac{3\sqrt{3}}{4\pi \left(5+2\sqrt{6}\right)}a^{2} \\
v_{c} &=& \frac{3\sqrt{2}+2\sqrt{3}}{3a} \\
T_{c} &=& \frac{a}{2\pi \left(5+2\sqrt{6}\right)} ~. \label{ax3}
\end{eqnarray}
It is {unusual} that the critical pressure is negative. 
{Being more usual} the critical constants are {dependent}
on {the} Rindler parameter. 
{Then the} critical ratio is given by 
\begin{eqnarray}
\rho_{c} &=& -\frac{\sqrt{3}}{2}\left(3\sqrt{2}+2\sqrt{3}\right) ~. \label{ax4}
\end{eqnarray}
It is also {a peculiar} result that the critical constant is negative.

\section*{Acknowledgements}
 I am  grateful to  Dr. Daniel Grumiler for {his most helpful} email correspondence.


\end{document}